\documentclass[prx,reprint,superscriptaddress,longbibliography,floatfix,nofootinbib]{revtex4-1}
\usepackage{amsmath}
\usepackage{mathtools, nccmath}
\usepackage{setspace}
\usepackage{hyperref}
\usepackage{outlines}
\usepackage{fancyvrb}
\usepackage{mathtools}
\usepackage{hyperref}
\usepackage{braket}
\usepackage{comment}
\usepackage{bbold}
\usepackage{amssymb}
\usepackage{graphicx}
\usepackage[caption=false]{subfig}
\usepackage[usenames, dvipsnames]{color}
\usepackage{soul}
\usepackage{tensor}
\usepackage{amsthm}
\usepackage[caption=false]{subfig}
\usepackage{float}
\makeatletter
\let\newfloat\newfloat@ltx
\makeatother
\usepackage{algorithm,algpseudocode}
\usepackage{algcompatible}

\newtheorem{theorem}{Theorem} 
\DeclarePairedDelimiter{\nint}\lfloor\rceil
\begin{document}
 
\title{Memory-assisted decoder for approximate Gottesman-Kitaev-Preskill codes}

\author{Kwok Ho Wan}
\email[Corresponding author: ]{kwok.wan14@imperial.ac.uk}
\affiliation{Mathematical Physics, Department of Mathematics,
  Imperial College London, London, SW7 2AZ, UK}
  \affiliation{QOLS, Blackett Laboratory, Imperial College London, London, SW7 2AZ, UK}

\author{Alex Neville}
  \affiliation{QOLS, Blackett Laboratory, Imperial College London, London, SW7 2AZ, UK}

\author{Steve Kolthammer} 
  \affiliation{QOLS, Blackett Laboratory, Imperial College London, London, SW7 2AZ, UK}
\date{\today}

\begin{abstract}
We propose a quantum error correction protocol for continuous-variable finite-energy, approximate Gottesman-Kitaev-Preskill (GKP) states undergoing small Gaussian random displacement errors, based on the scheme of Glancy and Knill [Phys. Rev. A {\bf 73}, 012325 (2006)].
We show that combining multiple rounds of error-syndrome extraction with Bayesian estimation offers enhanced protection of GKP-encoded qubits over comparible single-round approaches.
Furthermore, we show that the expected total displacement error incurred in multiple rounds of error followed by syndrome extraction is bounded by $2\sqrt{\pi}$.
By recompiling the syndrome-extraction circuits, we show that all squeezing operations can be subsumed into auxiliary state preparation, reducing them to beamsplitter transformations and quadrature measurements.
\end{abstract}


\maketitle

\section{Introduction} Encoding and manipulating quantum information in continuous variable (CV) systems~\cite{Lloyd1999, Braunstein2005, weedbrook2012gaussian} is a promising route to realising a useful quantum computing device.
Large scale CV cluster states can be generated on demand~\cite{furu_1_mil}, and fast, high-quality one- and two-qubit Clifford gates are deterministically available~\cite{furusawa2011quantum}.
Fault tolerant, measurement-based quantum computation is possible using CV cluster states along with CV measurements and non-Gaussian state injection ~\cite{Menicucci2014,2019arXiv190300012B}, though the levels of squeezing required for fault tolerance are beyond the reach of current experiments~\cite{Menicucci2014,PhysRevX.8.021054,2019arXiv190803579N,Fluhmann2019,2019arXiv190712487C}.

A leading approach to CV quantum computation, proposed by Gottesman, Kitaev and Preskill~\cite{PhysRevA.64.012310}, is based on the idea of encoding a qubit within an (infinite dimensional) oscillator.
Ideal codeword wavefunctions within this paradigm correspond to infinite-energy Dirac combs, and are commonly referred to as GKP states.
In practice, this ideal wavefunction is replaced by a finite-energy approximation, such as a comb of narrow Gaussian peaks modulated by a broad Gaussian envelope.

The appeal of GKP-encoded qubits is that they possess both an intrinsic robustness to physically motivated error channels and natural schemes for error syndrome extraction and correction.
An initial proposal for GKP error correction was based on CV stabilizer generalisations of the Steane circuits~\cite{PhysRevLett.78.2252} for error syndrome extraction.
Subsequently, Glancy and Knill proposed a different method based on a beamsplitter transformation~\cite{PhysRevA.73.012325}, which will provide the basis for the scheme detailed in this paper.

We define approximate GKP codewords with width $\vec{\Delta} = (\Delta,\kappa)$ as
\begin{equation}
   \psi_{\mu}^{\vec{\Delta}} (x) \propto \sum_{s \in \mathbb{Z}} G_{\frac{1}{\kappa}} [(2s+\mu)\sqrt{\pi} \hspace{0.04cm}] \hspace{0.05cm} G_{\Delta}[x-(2s+\mu)\sqrt{\pi} \hspace{0.04cm}] 
\end{equation}
where $G_{\Sigma}(z) = \exp{(-\frac{z^2}{2\Sigma^2})}$ and $\mu \in \{0,1\}$ defines the logical basis states.
Informally, for the logical 0 (1) state we have a superposition of Gaussians of width $\Delta$ centered at even (odd) values of $\sqrt{\pi}$, with an overall Gaussian envelope of width $1/\kappa$.
Better approximations to the ideal GKP state are achieved with smaller values of $\Delta$ and $\kappa$, although these also correspond to larger average energy and an apparent increase in experimental difficulty.

Recently, Albert et al.~\cite{2017arXiv170805010A} showed that the GKP code outperforms a number of other bosonic codes when states are exposed to amplitude damping and Gaussian random displacement errors. We will consider the latter, and write such an error acting on the state $\hat{\rho}$ as
\begin{equation}
\mathcal{E}_{\sigma_0}(\hat{\rho}) = \frac{1}{2\pi\sigma_0^2}\displaystyle{\int\!\!\!\int \displaylimits_{\hspace{-0.2cm}\alpha\in\mathbb{C}}} \!\! \hspace{0.2cm} \text{d}^{2}\alpha \hspace{0.2cm} G_{\sigma_0}(|\alpha|) \hspace{0.05cm} \hat{D}(\alpha) \hspace{0.05cm} \hat{\rho} \hspace{0.05cm} \hat{D}(\alpha)^{\dagger} \ , 
\end{equation}
where the operator $\hat{D}(\alpha)$ shifts the state in phase space by Re\{$\alpha$\}, Im\{$\alpha$\} in the q- and p-quadratures respectively, and the width $\sigma_0$ quantifies the extent of the error.    Notably, this error model describes amplitude damping that is preceded by an offsetting pre-amplification \cite{8482307}, and it is therefore highly relevant to many experimental platforms. Despite this potential, explicit error correction protocols for accessible approximate states are currently lacking.

In this manuscript we present a new decoder for the GKP code undergoing Gaussian random displacement errors, based on the Glancy and Knill error recovery scheme and Bayesian estimation.
In particular, we extend the scheme to enable enhanced error estimation using multiple syndrome extractions, enabling improved error supression which we show to be useful in extending the lifetime of states with a mean number of bosons as low as ten.
We find that many rounds of syndrome extraction without active corrective displacement causes the qubit to drift in phase space by at most approximately $2\sqrt{\pi}$ in each quadrature.
Additionally, we recompile the syndrome extraction circuit in an experimentally friendly way, such that squeezing need only be applied to auxiliary states which can be prepared offline.

\section{Syndrome extraction}   
\begin{figure}[!t]
\center
	\includegraphics[width=0.8\linewidth]{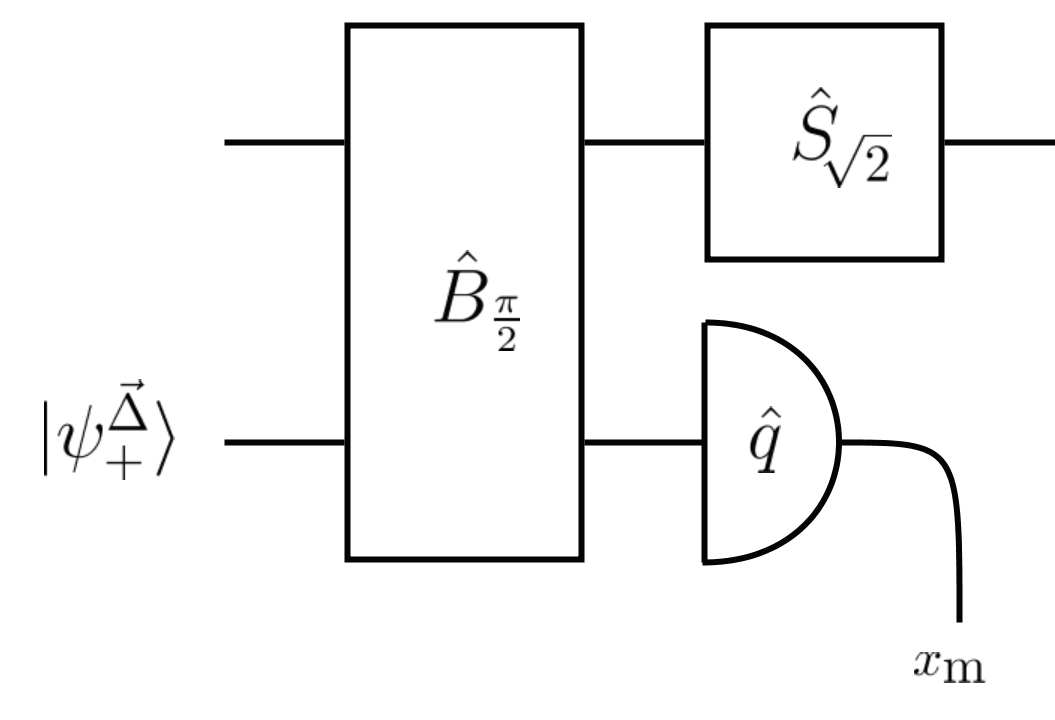}
	\vspace{-0.4cm}
	\caption{ \label{fig:GK_circuit_q}The q-quadrature error-syndrome extraction circuit (q-SE), as proposed by Glancy and Knill~\cite{PhysRevA.73.012325}. The GKP qubit and an auxiliary state are input in the top and bottom modes respectively. Following beamsplitting and squeezing operations, the error syndrome $x_\text{m}$ is generated by a q-quadrature measurement on the auxiliary mode.}
\end{figure}

The GKP syndrome extraction scheme of Glancy and Knill ~\cite{PhysRevA.73.012325} can be broken down into two sequential circuits: q-SE and p-SE, which extract the error syndromes in the q- and p-quadrature respectively (see Fig.~\ref{fig:GK_circuit_q}).

We begin our analysis with an arbitrary input qubit wavefunction $Q^{\vec{\Delta}}(x)$ that has undergone an unknown displacement error $(u,v)$

\begin{equation}
Q^{\vec{\Delta}}(x) \xrightarrow[(u,v)]{\text{error}} \text{e}^{ivx} Q^{\vec{\Delta}}(x-u) \ .
\end{equation}
This corrupted qubit is input into the top mode of the q-SE circuit, while an auxiliary GKP state $\ket{\psi^{\vec{\Delta}}_+} \propto \ket{\psi^{\vec{\Delta}}_{0}}+\ket{\psi^{\vec{\Delta}}_{1}}$ is input into the bottom mode. The action of the q-SE circuit can be visualised in terms of the two-mode q-quadrature wavefunction. The beamsplitter $\hat{B}_{\frac{\pi}{2}}$ causes an anticlockwise rotation by $45^{\circ}$ in the joint quadrature space, and the subsequent squeezer $\hat{S}_{\sqrt{2}}$ scales the top-mode quadrature by $\sqrt{2}$. The error syndrome is then generated by a q-quadrature measurement of the auxiliary mode.

The p-SE circuit proceeds similarly, although there are subtle differences beyond a change of variables $q \rightarrow p$, and squeezing in the conjugate direction ($\hat{S}_{\sqrt{2}}\rightarrow \hat{S}^{\dagger}_{\sqrt{2}}$).
Indeed, we must also change the auxiliary state $\ket{\psi_+^{\vec{\Delta}}} \rightarrow \ket{\psi_0^{\vec{\Delta}^\prime}}$ where $\vec{\Delta}^\prime = (\Delta/\sqrt{2}, \kappa\sqrt{2})$---a departure from the proposal of Glancy and Knill, which we found necessary for sequential q-SE and p-SE circuits to be applied to approximate GKP states (see Appendix\ \ref{appendix-single-step} for more details).

Following sequential q-SE and p-SE circuits, the qubit has transformed according to
\begin{equation}
\label{eq:approx_transformation}
\text{e}^{ivx} Q^{\vec{\Delta}}(x-u)\xrightarrow[\text{SE}]{\text{q-,p-}} \text{e}^{i\theta(p_\text{m},v)x} Q^{\vec{\Delta}}\big(x - \theta(x_\text{m},u)\big) \
\end{equation}
with a high fidelity (see Appendix\ \ref{appendix-success-prob}) when the displacement is small.
Here, $\theta(x_\text{m},u)$ is given by
\begin{equation}
    \theta(x_\text{m},u) = \frac{u}{2}-f^{*}_\text{step}(x_m)
\label{eq:single-step-theta}
\end{equation}
where $f^{*}_\text{step}(x_m)$ is the modified modular division by 4 function defined in Appendix\ \ref{appendix-defs}, and $\theta(p_\text{m},v)$ is defined similarly.
These quantities can be interpreted as the total displacement experienced by the qubit: an \emph{unknown}, random part $\frac{u}{2}$ remaining from the error channel and a \emph{known} part $f^{*}_\text{step}(x_m)$ introduced by the measurement.

\begin{figure*}
\captionsetup[subfigure]{position=top, labelfont=bf,textfont=normalfont,singlelinecheck=off,justification=raggedright}
\center
\vspace{-0.3cm}
\label{fig:mem}
\subfloat[]{\includegraphics[width=0.8\linewidth]{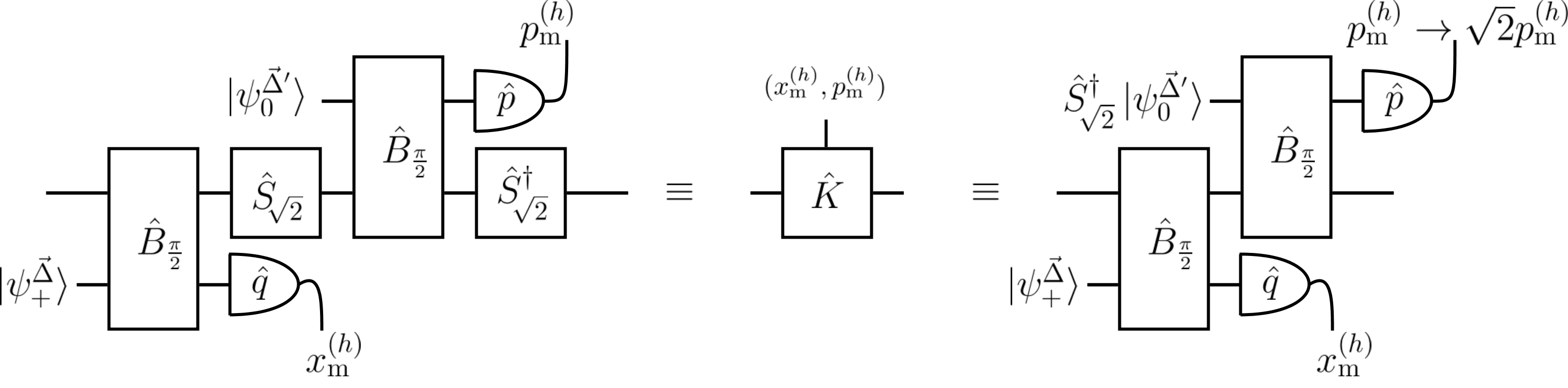}\label{fig:memory_module}}\\ 
\vspace{-0.3cm}
\subfloat[]{\includegraphics[width=0.8\linewidth]{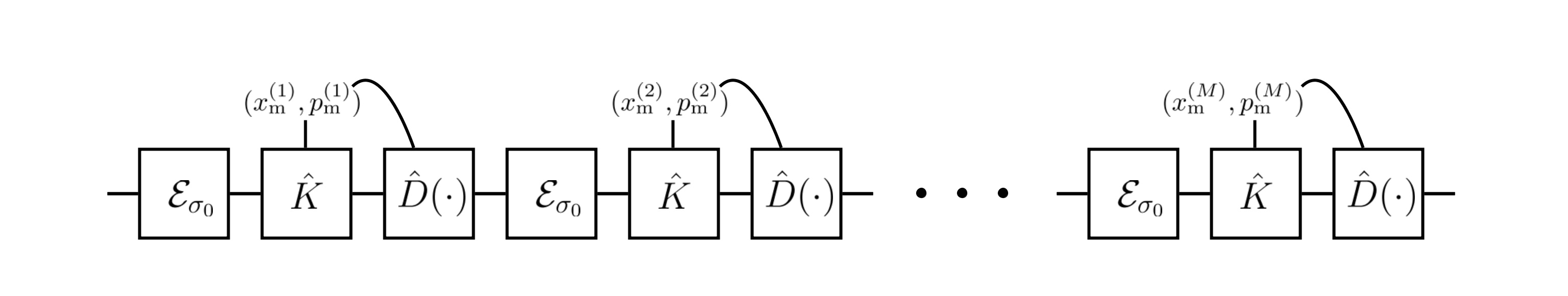}\label{fig:memoryless}}\\
\vspace{-0.3cm}
\subfloat[]{\includegraphics[width=0.8\linewidth]{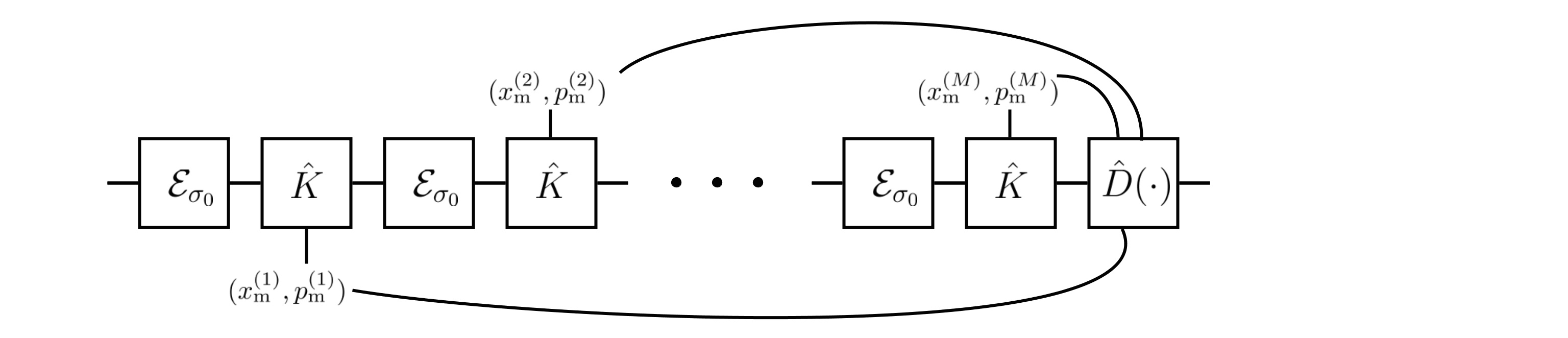}\label{fig:memory}}
\vspace{-0.3cm}
\caption{\label{fig:big} a) The standard q- and p-quadrature syndrome extraction circuits (left), which feature squeezing operations on the qubit mode, can be represented by a measurement dependent Kraus operator $\hat{K}$ (centre), and recompiled such that squeezing is applied to the p- auxiliary state, with a reinterpreted p-quadrature measurement (right).
b) Memoryless error correction---syndrome measurement results are used to inform a corrective displacement at each round, and no information is carried forward to future rounds. c) Memory-assisted error correction---no active corrective shift is performed after each syndrome measurement. Instead, all the syndrome measurement results are used together to decode and perform a single corrective displacement after $M$ rounds.}
\end{figure*}

While an error correcting procedure could involve estimation of the total displacement $\theta(x_m,u)$ and immediately applying a corrective displacement based on this estimate, we note that allowing an uncorrected qubit to undergo a further syndrome extraction process only results in another displaced version of the input qubit. With this in mind, our decoder uses measurement information from multiple rounds of errors and syndrome extraction without intermediate correction to estimate a single corrective displacement to apply. This memory-assisted approach is summarised in Fig.~\ref{fig:big}, alongside a contrasting memoryless approach which applies a correction after each syndrome extraction and then forgets about it.

We now proceed by extending our analysis from one round to multiple rounds of syndrome extraction.
Say that after the $h^{\text{th}}$ round of syndrome extraction we have histories of q-quadrature syndrome measurements $\vec{x}_\mathrm{m}=(x_\mathrm{m}^{(1)},\dotsc ,x_\mathrm{m}^{(h)})$ and displacement errors $\vec{u}=(u_1,\dotsc ,u_h)$, and similarly p-quadrature values $\vec{p}_\mathrm{m}=(p_\mathrm{m}^{(1)},\dotsc ,p_\mathrm{m}^{(h)})$ and $\vec{v}=(v_1,\dotsc ,v_h)$ for the p-quadrature.
If each displacement error is small, then  the input qubit is transformed according to
\begin{equation}
Q^{\vec{\Delta}}(x) \, \xrightarrow[]{} \, \text{e}^{i\theta_h(\vec{p}_\mathrm{m},\vec{v})x} Q^{\vec{\Delta}}\big(x-\theta_h(\vec{x}_\mathrm{m},\vec{u})\big).
\label{eq:multistep-approx-transformation}
\end{equation}
with high fidelity (again, see Appendix\ \ref{appendix-success-prob}). Focusing on the q-quadrature: the total shift in q is given by
\begin{equation}
\begin{split}
\theta_h\big(\vec{x}_\mathrm{m},\vec{u}\big) &= \sum_{j=1}^{h}\Big[\frac{u_j}{2^{h-j+1}}\Big] -
\sum_{k=1}^{h}\Big[\frac{f_\text{step}^{*}\big(\mathcal{X}_\text{m}^{(k)}(\vec{x}_\mathrm{m})\big)}{2^{h-k}}\Big] \\
&= \theta_h^{\mathrm{err}}(\vec{u})-\theta_h^{\mathrm{step}}(\vec{x}_\mathrm{m}) \ ,
\label{eq:reference_frame}
\end{split}
\end{equation}
where $\mathcal{X}^{(k)}_{\text{m}}$ can be interpreted as the measurement outcome $x^{(k)}_{\text{m}}$ transformed to account for previous measurement-induced shifts, and is given explicitly by
\begin{equation}
\mathcal{X}^{(h)}_{\text{m}}(\vec{x}_\mathrm{m}) = x^{(h)}_{\text{m}} + \frac{1}{\sqrt{2}}\sum_{j=1}^{k-1}\Big[\frac{f_\text{step}^{*}\big(\mathcal{X}_\text{m}^{(j)}(\vec{x}_\mathrm{m})\big)}{2^{k-j-1}}\Big]
\end{equation}
for $h>1$, and $\mathcal{X}^{(1)}_{\text{m}} = x^{(1)}_{\text{m}}$. In analogy with the single round case above, we have an unknown contribution to the total displacement $\theta_h^{\mathrm{err}}(\vec{u})$ together with a known, measurement-induced contribution $\theta_h^{\mathrm{step}}(\vec{x}_\mathrm{m})$.

Without active corrective shifts at each round, the distance a GKP qubit drifts in phase space after $h$ rounds is bounded above as: $\Big|\theta(\vec{x}_\text{m},\vec{u})\Big| < 2\sqrt{\pi} \ (1-2^{-h}) + |X|$, where $X \sim \mathcal{N}\Big(0,\sigma_0^2\frac{(1-4^{-h})}{3}\Big)$ is a Gaussian random variable with mean $0$ and variance $\sigma_0^2\frac{(1-4^{-h})}{3}$ (see Appendix\ \ref{appendix-multi-step} for proof). It follows that the expected value for total displacement error in each quadrature after any number of syndrome extraction rounds is bounded above by approximately $2\sqrt{\pi}$.


In Fig.~\ref{fig:memory_module}, we identify a practical simplification to the combined q-SE, p-SE circuit that moves all squeezing operations offline onto auxiliary state preparation. To do so, we use a modified p-SE auxiliary state, $\hat{S}^{\dagger}_{\sqrt{2}}\ket{\psi_0^{\vec{\Delta}'}}$, and reinterpret the p-quadrature measurement ($p_{\text{m}}\rightarrow\sqrt{2}p_\text{m}$) (see Appendix\ \ref{appendix-offline-squeezing} for proof).

\section{Decoder (Bayesian estimation)}
\label{section-decoder}
We now show how to use the q- and p-SE measurement outcomes to estimate the final displacements when the loss channel is well characterised (i.e. $\sigma_0$ is known).
For convenience, we set $\kappa = \Delta$ in this section, although it is straightforward to extend the results to the more general case.


In the case of a single round of q-SE and p-SE application, the probability of measuring $x_\text{m}$ given a shift error $u$ can be approximated as $\mathbb{P}(x_\text{m}|u) \propto \psi^{\vec{\Delta}}_{+}(\sqrt{2}x_\text{m} - u)$, which we note is independent of both $v$ and $p_m$.
By using a prior probability density function (PDF) corresponding to our characterised error channel and applying Bayes' theorem, we obtain a posterior PDF
\begin{equation}
\mathbb{P}(u|x_\text{m}) \propto \psi^{\vec{\Delta}}_{+}(\sqrt{2}x_\text{m} - u)\text{e}^{-\frac{u^2}{2\sigma_0^2}}.
\end{equation}
Assuming that $\vec{\Delta}$ and $\sigma_0$ are both small compared to $\sqrt{\pi}$, this posterior is well approximated by
\begin{equation}
    \mathbb{P}(u|x_\text{m}) \approx \mathcal{N}\left(\frac{\sqrt{2} \sigma_0^2 x_\text{m} - \sqrt{\pi} \sigma_0^2  \nint*{\frac{\sqrt{2}x_\text{m}}{\sqrt{\pi}}}}{\Delta^2 + \sigma_0^2}, \frac{\Delta^2\sigma_0^2}{\Delta^2 + \sigma_0^2}\right) 
\end{equation}
and we take our estimate $\tilde{u}$ of the shift error $u$ to be the mean of this Gaussian.
Explicitly,
\begin{equation}
    \tilde{u} = \frac{\sqrt{2} \sigma_0^2 x_\text{m} - \sqrt{\pi} \sigma_0^2  \nint*{\frac{\sqrt{2}x_\text{m}}{\sqrt{\pi}}}}{\Delta^2 + \sigma_0^2},
    \label{eq:single-step-u-est}
\end{equation}
which is a good approximation of the minimum mean square error (MMSE) estimator for $u$.
The estimate for the displacement required to counteract the acquired error is therefore $\theta(x_\text{m},\tilde{u})$, from combining equations\ \eqref{eq:single-step-theta} and~\eqref{eq:single-step-u-est}.

The corresponding p-quadrature conditional probabilities can be obtained by substituting 
$(u,x_{\text{m}},\Delta) \rightarrow (v,p_\text{m}, 2\Delta)$, where the asymmetry between the q- and p-quadrature originates from the differing widths of the auxiliary states.

For multiple rounds, the estimation of the cumulative unknown displacement caused by the vector $\vec{u}$ is achieved in the same spirit as estimating the parameter $u$ in the single round case, except with a messier-looking (although, still efficiently computable) estimator.

The probability of obtaining the q-quadrature measurement outcome $x_{\text{m}}^{(h)}$ in the $h^{\text{th}}$ round, given the measurement values obtained in the previous $(h-1)$ rounds and displacement shifts in all $h$ rounds, is  $\mathbb{P}(x^{(h)}_{\text{m}}|u_1,...,u_h,x^{(1)}_{\text{m}},...,x^{(h-1)}_{\text{m}}) \propto \psi^{\vec{\Delta}}_{+}(\sqrt{2}x^{(h)}_{\text{m}}-\mathcal{U}_h)$.
Here, $\mathcal{U}_h$ is the shift contributed in the $h^\text{th}$ round by the error channel, transformed to account for all previous steps (see Appendix\ \ref{appendix-defs} for an explicit definition).
We now apply Bayes' theorem with Gaussian prior PDFs for the error in each of $M$ rounds, resulting in the posterior PDF
\begin{equation}
    \mathbb{P}^{(\text{q})}_{M}\left(\vec{u} \big| \vec{x}_\text{m} \right) \propto \prod_{h=1}^{M}\psi^{\vec{\Delta}}_{+}\left(\sqrt{2}x^{(h)}_{\text{m}}-\mathcal{U}_h\right) \cdot \hspace{0.05cm} G_{\sigma_0}\left(u_h\right) \ , 
    \label{eq:PUXM}
\end{equation}

Assuming, again, that $\sigma_0$ and $\Delta$ are small compared to $\sqrt{\pi}$, we show that the posterior $\mathbb{P}^{(\text{q})}_{M}(\vec{u}|\vec{x}_\text{m})$ is well approximated by a multivariate Gaussian $\mathcal{N}(\vec{\tilde{u}},\Sigma)$ with mean vector $\vec{\tilde{u}}$ and covariance matrix $\Sigma$, in analogy with the single round case.
The quantity that we actually wish to estimate is $\theta_M^{\mathrm{err}}(\vec{u})$ from equation\ \eqref{eq:reference_frame}, which is a linear combination of elements from $\vec{u}$. The corresponding PDF is given by

\begin{equation}
\mathbb{P}\left(\theta_M^{\mathrm{err}}(\vec{u}) \big| \vec{x}_{\text{m}}\right) \approx \mathcal{N}\left(\vec{a}\cdot\vec{\tilde{u}} \ , \ \vec{a}^{\hspace{0.05cm}\text{T}}\hspace{-0.05cm} 
   \cdot\Sigma\cdot
    \vec{a} \right) \ ,
\end{equation}
\noindent with $a_{k} = 2^{-(M+1-k)}$ and
\begin{equation}
    \begin{split}
    \tilde{u}_{k} = \Big(\frac{\sigma_0}{\Delta}\Big)^2 \Bigg\{& 2^{k}\sum_{j=1}^{M}\frac{F_j}{2^j} \\ 
    & - \Big(\frac{\sigma_0}{\Delta}\Big)^2\sum_{h=1}^{M}\Big[\Big(\hspace{-0.2cm}\sum_{n=\text{m}_{k,h}}^{M}\hspace{-0.2cm}\frac{2^{k+h}}{4^n}\Big)\Big(\sum_{j=h}^{M}\frac{F_j}{2^j}\Big)\Big]2^{h}\Bigg\} \ , 
    \end{split}
    \label{eq:u_tilde_k}
\end{equation}
\noindent to order $\big(\frac{\sigma_0}{\Delta}\big)^4$, where $\text{m}_{k,h} = \text{max}\{k,h\}$ and $F_h = \sqrt{2}\mathcal{X}_\text{m}^{(h)}-\sqrt{\pi}\nint*{\sqrt{2}\mathcal{X}_\text{m}^{(h)}/\sqrt{\pi}}$.

As in the single round case, the posterior mean is used as the MMSE estimator for the displacement error.
We see that the posterior is well approximated by a Gaussian with a mean $\vec{a}\cdot\vec{\tilde{u}}$ which can be calculated directly from syndrome measurement results. An explicit approach to this calculation is presented as an algorithm in Appendix\ \ref{appendix-algorithm}.

Also of interest is the variance of this Gaussian, $\vec{a}^{\hspace{0.05cm}\text{T}}\hspace{-0.05cm} \cdot\Sigma\cdot \vec{a}$, as it corresponds to the uncertainty in our estimation.
In Appendix\ \ref{appendix-multi-step} we show that, if $\frac{\sigma_0}{\Delta} < \frac{1}{2}$ (which is true for all $M>1$), the variance converges to
\begin{equation}
\label{eq:var_q_multiple_steps1}
\begin{split}
V_\text{q}(M) & = \frac{\sigma_{0}^2}{3}\Big[(1-4^{-M}) \\ 
& + \Big(\frac{\sigma_0}{\Delta}\Big)^{2}\frac{4}{9}\big(4^{-2M} + 3(1+2M)4^{-M} -4\big) \Big],
\end{split}
\end{equation}
\noindent neglecting terms within the bracket of order $\big( \frac{\sigma_0}{\Delta} \big)^4$ and higher. Note that $V_\text{q} \rightarrow \frac{\sigma_{0}^2}{3}$ very quickly as $M$ grows, given that $\frac{\sigma_0}{\Delta} \ll 1$. 

\begin{figure}[!t]
\center
	\includegraphics[width=1\linewidth]{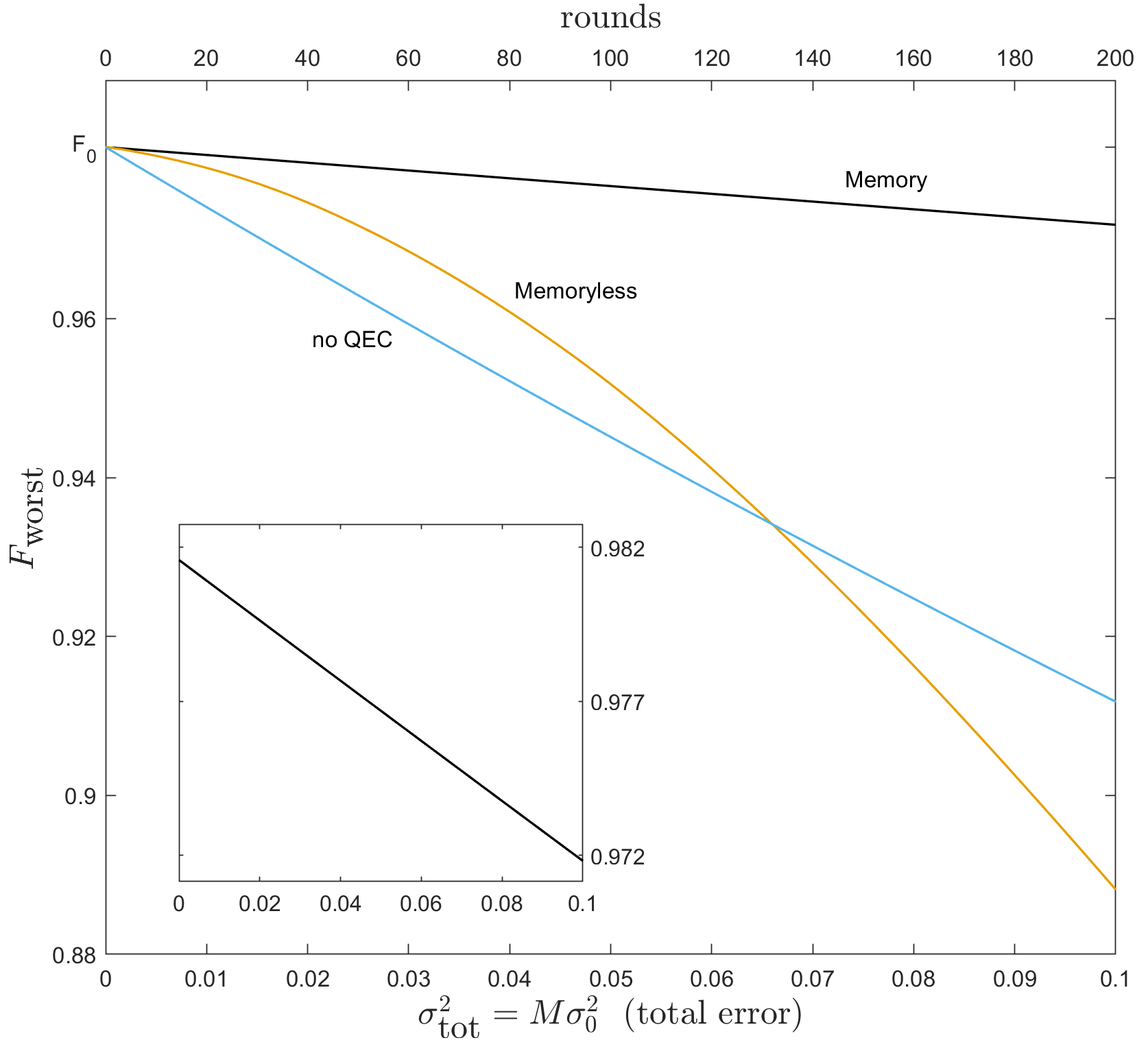}
	\caption{\label{fig:n_10_multiple_step} Qubit fidelity achieved by the memory-assisted decoder (black), memoryless decoder (yellow) and no QEC (aqua). A Gaussian error channel of width $\sigma_0^2  = 0.0005$ is applied in each of the $M$ rounds of syndrome extraction to a GKP-state with width $\Delta = \kappa = 0.22$. The initial fidelity $F_0 = 0.981$, and the inset plot shows how the black curve decreases over a smaller range.}
\end{figure}

\section{Numerical results}   The aim of a quantum error correction (QEC) procedure is to protect logical quantum information. 
Pantaleoni et al.~recently reported a method to extract logical information from approximate GKP states~\cite{2019arXiv190708210P}. In essence, this reduces a density matrix function describing a CV state to a qubit density matrix ($\hat{\rho}^{\text{CV}}\rightarrow \hat{\rho}^{\text{qubit}}$). 
Using this, fidelities between qubit density matrices at the input and output of an error correction procedure can be computed in order to benchmark its performance.

In Fig.~\ref{fig:n_10_multiple_step} we present the results of numerical simulations for the specific case of a GKP code with $\Delta \approx 0.22$ (corresponding to an average number of bosons $\bar{n}\approx 10$). 
For varying numbers of rounds of error channel application followed by syndrome extraction, we benchmark error correction with our memory-assisted decoder against density matrix function simulations of the memoryless decoder and no error correction at all.

We observe that error correction is improved by using our memory-assisted decoder instead of the memoryless decoder ---
a finding similar to that of Vuillot et al.~\cite{PhysRevA.99.032344} and Noh et al.~\cite{2019arXiv190803579N} for Steane-based GKP syndrome extraction. Note that $\Delta$ is large enough that the memoryless decoder performs worse than no QEC once a few hundred rounds are considered. We also see that the quality of the memory-assisted error corrected state is higher than the uncorrected state, despite the state containing a low average number of bosons. Furthermore, the fidelity of the input state is approximately preserved through application of the memory-assisted error correction scheme.

\section{Summary and outlook}
We have described an explicit protocol for GKP quantum error correction that provides improved protection from Gaussian displacement errors with reduced experimental requirements. Notably, we have shown that GKP states undergo a total displacement (and therefore require a correction) that is approximately bounded by $2\sqrt{\pi}$ in each quadrature, after multiple rounds of error syndrome extraction. A memory-assisted decoder based on Bayesian estimation was developed to specify the near-optimal corrective displacement, and numerical simulations have shown that this results in significantly improved error correction compared to a memoryless decoder when applied to an approximate GKP state with a small average number of bosons.

Recent experiments have demonstrated GKP states encoded in oscillations of trapped ions~\cite{Fluhmann2019} and microwave fields of superconducting resonators~\cite{2019arXiv190712487C}. In the latter case, states are shown with widths of $(\Delta, \kappa) = (0.16, 0.32)$. 
As these values are similar to those we have studied here, this suggests that our decoder can be exploited by near-term experiments as part of an error correction procedure. Additionally, the Glancy-Knill syndrome extraction circuit at the core of our method may be better suited to realistic devices than the more studied SUM gate syndrome extraction, since the latter requires more gates, and therefore more physical components, to implement~\cite{TBMSK2019}.

The error model that we have considered in this work  also pertains to errors that arise in teleporting a GKP state along a CV cluster state with finite squeezing~\cite{PhysRevA.79.062318}.
In this situation, the finite squeezing of the cluster state induces a displacement error. 
We anticipate that it will therefore be possible to apply our error correction scheme between nodes on a CV cluster state graph, and thereby lower the squeezing threshold for universal fault tolerant quantum computation using CV cluster states and GKP state injection.
\section*{Data availability statement}
This is a theoretical paper and there is no experimental data available beyond the numerical simulation data described in the paper. Kwok Ho Wan performed all numerical experiments, all the authors contributed to the manuscript and the project was supervised by Alex Neville and Steve Kolthammer.
\section*{Acknowledgements}
We acknowledge discussions with Luca Cocconi, Jacob Hastrup, Hl\'er Kristj\'ansson, Y. H. Chang, Frederic Sauvage, Jonathan Conrad, Christophe Vuillot, Kyungjoo Noh, Giacomo Pantaleoni, Myungshik Kim and Hyukjoon Kwon. Kwok Ho Wan is funded by the President's PhD Scholarship of Imperial College London.
Steve Kolthammer acknowledges support from the EPSRC grant EP/T001062/1.

\bibliography{main.bib}

\begin{thebibliography}{20}%
\makeatletter
\providecommand \@ifxundefined [1]{%
 \@ifx{#1\undefined}
}%
\providecommand \@ifnum [1]{%
 \ifnum #1\expandafter \@firstoftwo
 \else \expandafter \@secondoftwo
 \fi
}%
\providecommand \@ifx [1]{%
 \ifx #1\expandafter \@firstoftwo
 \else \expandafter \@secondoftwo
 \fi
}%
\providecommand \natexlab [1]{#1}%
\providecommand \enquote  [1]{``#1''}%
\providecommand \bibnamefont  [1]{#1}%
\providecommand \bibfnamefont [1]{#1}%
\providecommand \citenamefont [1]{#1}%
\providecommand \href@noop [0]{\@secondoftwo}%
\providecommand \href [0]{\begingroup \@sanitize@url \@href}%
\providecommand \@href[1]{\@@startlink{#1}\@@href}%
\providecommand \@@href[1]{\endgroup#1\@@endlink}%
\providecommand \@sanitize@url [0]{\catcode `\\12\catcode `\$12\catcode
  `\&12\catcode `\#12\catcode `\^12\catcode `\_12\catcode `\%12\relax}%
\providecommand \@@startlink[1]{}%
\providecommand \@@endlink[0]{}%
\providecommand \url  [0]{\begingroup\@sanitize@url \@url }%
\providecommand \@url [1]{\endgroup\@href {#1}{\urlprefix }}%
\providecommand \urlprefix  [0]{URL }%
\providecommand \Eprint [0]{\href }%
\providecommand \doibase [0]{http://dx.doi.org/}%
\providecommand \selectlanguage [0]{\@gobble}%
\providecommand \bibinfo  [0]{\@secondoftwo}%
\providecommand \bibfield  [0]{\@secondoftwo}%
\providecommand \translation [1]{[#1]}%
\providecommand \BibitemOpen [0]{}%
\providecommand \bibitemStop [0]{}%
\providecommand \bibitemNoStop [0]{.\EOS\space}%
\providecommand \EOS [0]{\spacefactor3000\relax}%
\providecommand \BibitemShut  [1]{\csname bibitem#1\endcsname}%
\let\auto@bib@innerbib\@empty
\bibitem [{\citenamefont {Lloyd}\ and\ \citenamefont
  {Braunstein}(1999)}]{Lloyd1999}%
  \BibitemOpen
  \bibfield  {author} {\bibinfo {author} {\bibfnamefont {Seth}\ \bibnamefont
  {Lloyd}}\ and\ \bibinfo {author} {\bibfnamefont {Samuel~L.}\ \bibnamefont
  {Braunstein}},\ }\bibfield  {title} {\enquote {\bibinfo {title} {Quantum
  computation over continuous variables},}\ }\href {\doibase
  10.1103/PhysRevLett.82.1784} {\bibfield  {journal} {\bibinfo  {journal}
  {Phys. Rev. Lett.}\ }\textbf {\bibinfo {volume} {82}},\ \bibinfo {pages}
  {1784--1787} (\bibinfo {year} {1999})}\BibitemShut {NoStop}%
\bibitem [{\citenamefont {Braunstein}\ and\ \citenamefont {van
  Loock}(2005)}]{Braunstein2005}%
  \BibitemOpen
  \bibfield  {author} {\bibinfo {author} {\bibfnamefont {Samuel~L.}\
  \bibnamefont {Braunstein}}\ and\ \bibinfo {author} {\bibfnamefont {Peter}\
  \bibnamefont {van Loock}},\ }\bibfield  {title} {\enquote {\bibinfo {title}
  {Quantum information with continuous variables},}\ }\href {\doibase
  10.1103/RevModPhys.77.513} {\bibfield  {journal} {\bibinfo  {journal} {Rev.
  Mod. Phys.}\ }\textbf {\bibinfo {volume} {77}},\ \bibinfo {pages} {513--577}
  (\bibinfo {year} {2005})}\BibitemShut {NoStop}%
\bibitem [{\citenamefont {Weedbrook}\ \emph {et~al.}(2012)\citenamefont
  {Weedbrook}, \citenamefont {Pirandola}, \citenamefont
  {Garc{\'\i}a-Patr{\'o}n}, \citenamefont {Cerf}, \citenamefont {Ralph},
  \citenamefont {Shapiro},\ and\ \citenamefont
  {Lloyd}}]{weedbrook2012gaussian}%
  \BibitemOpen
  \bibfield  {author} {\bibinfo {author} {\bibfnamefont {Christian}\
  \bibnamefont {Weedbrook}}, \bibinfo {author} {\bibfnamefont {Stefano}\
  \bibnamefont {Pirandola}}, \bibinfo {author} {\bibfnamefont {Ra{\'u}l}\
  \bibnamefont {Garc{\'\i}a-Patr{\'o}n}}, \bibinfo {author} {\bibfnamefont
  {Nicolas~J}\ \bibnamefont {Cerf}}, \bibinfo {author} {\bibfnamefont
  {Timothy~C}\ \bibnamefont {Ralph}}, \bibinfo {author} {\bibfnamefont
  {Jeffrey~H}\ \bibnamefont {Shapiro}}, \ and\ \bibinfo {author} {\bibfnamefont
  {Seth}\ \bibnamefont {Lloyd}},\ }\bibfield  {title} {\enquote {\bibinfo
  {title} {Gaussian quantum information},}\ }\href@noop {} {\bibfield
  {journal} {\bibinfo  {journal} {Reviews of Modern Physics}\ }\textbf
  {\bibinfo {volume} {84}},\ \bibinfo {pages} {621} (\bibinfo {year}
  {2012})}\BibitemShut {NoStop}%
\bibitem [{\citenamefont {{Yoshikawa}}\ \emph {et~al.}(2016)\citenamefont
  {{Yoshikawa}}, \citenamefont {{Yokoyama}}, \citenamefont {{Kaji}},
  \citenamefont {{Sornphiphatphong}}, \citenamefont {{Shiozawa}}, \citenamefont
  {{Makino}},\ and\ \citenamefont {{Furusawa}}}]{furu_1_mil}%
  \BibitemOpen
  \bibfield  {author} {\bibinfo {author} {\bibfnamefont {Jun-ichi}\
  \bibnamefont {{Yoshikawa}}}, \bibinfo {author} {\bibfnamefont {Shota}\
  \bibnamefont {{Yokoyama}}}, \bibinfo {author} {\bibfnamefont {Toshiyuki}\
  \bibnamefont {{Kaji}}}, \bibinfo {author} {\bibfnamefont {Chanond}\
  \bibnamefont {{Sornphiphatphong}}}, \bibinfo {author} {\bibfnamefont
  {Yu}~\bibnamefont {{Shiozawa}}}, \bibinfo {author} {\bibfnamefont {Kenzo}\
  \bibnamefont {{Makino}}}, \ and\ \bibinfo {author} {\bibfnamefont {Akira}\
  \bibnamefont {{Furusawa}}},\ }\bibfield  {title} {\enquote {\bibinfo {title}
  {{Invited Article: Generation of one-million-mode continuous-variable cluster
  state by unlimited time-domain multiplexing}},}\ }\href {\doibase
  10.1063/1.4962732} {\bibfield  {journal} {\bibinfo  {journal} {APL
  Photonics}\ }\textbf {\bibinfo {volume} {1}},\ \bibinfo {eid} {060801}
  (\bibinfo {year} {2016})},\ \Eprint {http://arxiv.org/abs/1606.06688}
  {arXiv:1606.06688 [quant-ph]} \BibitemShut {NoStop}%
\bibitem [{\citenamefont {Furusawa}\ and\ \citenamefont {van
  Loock}(2011)}]{furusawa2011quantum}%
  \BibitemOpen
  \bibfield  {author} {\bibinfo {author} {\bibfnamefont {A.}~\bibnamefont
  {Furusawa}}\ and\ \bibinfo {author} {\bibfnamefont {P.}~\bibnamefont {van
  Loock}},\ }\href@noop {} {\emph {\bibinfo {title} {Quantum Teleportation and
  Entanglement, A Hybrid Approach to Optical Quantum Information Processing}}}\
  (\bibinfo  {publisher} {Wiley},\ \bibinfo {year} {2011})\BibitemShut
  {NoStop}%
\bibitem [{\citenamefont {Menicucci}(2014)}]{Menicucci2014}%
  \BibitemOpen
  \bibfield  {author} {\bibinfo {author} {\bibfnamefont {Nicolas~C.}\
  \bibnamefont {Menicucci}},\ }\bibfield  {title} {\enquote {\bibinfo {title}
  {Fault-tolerant measurement-based quantum computing with continuous-variable
  cluster states},}\ }\href {\doibase 10.1103/PhysRevLett.112.120504}
  {\bibfield  {journal} {\bibinfo  {journal} {Phys. Rev. Lett.}\ }\textbf
  {\bibinfo {volume} {112}},\ \bibinfo {pages} {120504} (\bibinfo {year}
  {2014})}\BibitemShut {NoStop}%
\bibitem [{\citenamefont {{Baragiola}}\ \emph {et~al.}(2019)\citenamefont
  {{Baragiola}}, \citenamefont {{Pantaleoni}}, \citenamefont {{Alexand er}},
  \citenamefont {{Karanjai}},\ and\ \citenamefont
  {{Menicucci}}}]{2019arXiv190300012B}%
  \BibitemOpen
  \bibfield  {author} {\bibinfo {author} {\bibfnamefont {Ben~Q.}\ \bibnamefont
  {{Baragiola}}}, \bibinfo {author} {\bibfnamefont {Giacomo}\ \bibnamefont
  {{Pantaleoni}}}, \bibinfo {author} {\bibfnamefont {Rafael~N.}\ \bibnamefont
  {{Alexand er}}}, \bibinfo {author} {\bibfnamefont {Angela}\ \bibnamefont
  {{Karanjai}}}, \ and\ \bibinfo {author} {\bibfnamefont {Nicolas~C.}\
  \bibnamefont {{Menicucci}}},\ }\bibfield  {title} {\enquote {\bibinfo {title}
  {{All-Gaussian universality and fault tolerance with the
  Gottesman-Kitaev-Preskill code}},}\ }\href@noop {} {\bibfield  {journal}
  {\bibinfo  {journal} {arXiv e-prints}\ ,\ \bibinfo {eid} {arXiv:1903.00012}}
  (\bibinfo {year} {2019})},\ \Eprint {http://arxiv.org/abs/1903.00012}
  {arXiv:1903.00012 [quant-ph]} \BibitemShut {NoStop}%
\bibitem [{\citenamefont {Fukui}\ \emph {et~al.}(2018)\citenamefont {Fukui},
  \citenamefont {Tomita}, \citenamefont {Okamoto},\ and\ \citenamefont
  {Fujii}}]{PhysRevX.8.021054}%
  \BibitemOpen
  \bibfield  {author} {\bibinfo {author} {\bibfnamefont {Kosuke}\ \bibnamefont
  {Fukui}}, \bibinfo {author} {\bibfnamefont {Akihisa}\ \bibnamefont {Tomita}},
  \bibinfo {author} {\bibfnamefont {Atsushi}\ \bibnamefont {Okamoto}}, \ and\
  \bibinfo {author} {\bibfnamefont {Keisuke}\ \bibnamefont {Fujii}},\
  }\bibfield  {title} {\enquote {\bibinfo {title} {High-threshold
  fault-tolerant quantum computation with analog quantum error correction},}\
  }\href {\doibase 10.1103/PhysRevX.8.021054} {\bibfield  {journal} {\bibinfo
  {journal} {Phys. Rev. X}\ }\textbf {\bibinfo {volume} {8}},\ \bibinfo {pages}
  {021054} (\bibinfo {year} {2018})}\BibitemShut {NoStop}%
\bibitem [{\citenamefont {{Noh}}\ and\ \citenamefont
  {{Chamberland}}(2019)}]{2019arXiv190803579N}%
  \BibitemOpen
  \bibfield  {author} {\bibinfo {author} {\bibfnamefont {Kyungjoo}\
  \bibnamefont {{Noh}}}\ and\ \bibinfo {author} {\bibfnamefont {Christopher}\
  \bibnamefont {{Chamberland}}},\ }\bibfield  {title} {\enquote {\bibinfo
  {title} {{Fault-tolerant bosonic quantum error correction with the
  surface-GKP code}},}\ }\href@noop {} {\bibfield  {journal} {\bibinfo
  {journal} {arXiv e-prints}\ ,\ \bibinfo {eid} {arXiv:1908.03579}} (\bibinfo
  {year} {2019})},\ \Eprint {http://arxiv.org/abs/1908.03579} {arXiv:1908.03579
  [quant-ph]} \BibitemShut {NoStop}%
\bibitem [{\citenamefont {Fl{\"u}hmann}\ \emph {et~al.}(2019)\citenamefont
  {Fl{\"u}hmann}, \citenamefont {Nguyen}, \citenamefont {Marinelli},
  \citenamefont {Negnevitsky}, \citenamefont {Mehta},\ and\ \citenamefont
  {Home}}]{Fluhmann2019}%
  \BibitemOpen
  \bibfield  {author} {\bibinfo {author} {\bibfnamefont {C.}~\bibnamefont
  {Fl{\"u}hmann}}, \bibinfo {author} {\bibfnamefont {T.~L.}\ \bibnamefont
  {Nguyen}}, \bibinfo {author} {\bibfnamefont {M.}~\bibnamefont {Marinelli}},
  \bibinfo {author} {\bibfnamefont {V.}~\bibnamefont {Negnevitsky}}, \bibinfo
  {author} {\bibfnamefont {K.}~\bibnamefont {Mehta}}, \ and\ \bibinfo {author}
  {\bibfnamefont {J.~P.}\ \bibnamefont {Home}},\ }\bibfield  {title} {\enquote
  {\bibinfo {title} {Encoding a qubit in a trapped-ion mechanical
  oscillator},}\ }\href {\doibase 10.1038/s41586-019-0960-6} {\bibfield
  {journal} {\bibinfo  {journal} {Nature}\ }\textbf {\bibinfo {volume} {566}},\
  \bibinfo {pages} {513--517} (\bibinfo {year} {2019})}\BibitemShut {NoStop}%
\bibitem [{\citenamefont {{Campagne-Ibarcq}}\ \emph {et~al.}(2019)\citenamefont
  {{Campagne-Ibarcq}}, \citenamefont {{Eickbusch}}, \citenamefont {{Touzard}},
  \citenamefont {{Zalys-Geller}}, \citenamefont {{Frattini}}, \citenamefont
  {{Sivak}}, \citenamefont {{Reinhold}}, \citenamefont {{Puri}}, \citenamefont
  {{Shankar}}, \citenamefont {{Schoelkopf}}, \citenamefont {{Frunzio}},
  \citenamefont {{Mirrahimi}},\ and\ \citenamefont
  {{Devoret}}}]{2019arXiv190712487C}%
  \BibitemOpen
  \bibfield  {author} {\bibinfo {author} {\bibfnamefont {P.}~\bibnamefont
  {{Campagne-Ibarcq}}}, \bibinfo {author} {\bibfnamefont {A.}~\bibnamefont
  {{Eickbusch}}}, \bibinfo {author} {\bibfnamefont {S.}~\bibnamefont
  {{Touzard}}}, \bibinfo {author} {\bibfnamefont {E.}~\bibnamefont
  {{Zalys-Geller}}}, \bibinfo {author} {\bibfnamefont {N.~E.}\ \bibnamefont
  {{Frattini}}}, \bibinfo {author} {\bibfnamefont {V.~V.}\ \bibnamefont
  {{Sivak}}}, \bibinfo {author} {\bibfnamefont {P.}~\bibnamefont {{Reinhold}}},
  \bibinfo {author} {\bibfnamefont {S.}~\bibnamefont {{Puri}}}, \bibinfo
  {author} {\bibfnamefont {S.}~\bibnamefont {{Shankar}}}, \bibinfo {author}
  {\bibfnamefont {R.~J.}\ \bibnamefont {{Schoelkopf}}}, \bibinfo {author}
  {\bibfnamefont {L.}~\bibnamefont {{Frunzio}}}, \bibinfo {author}
  {\bibfnamefont {M.}~\bibnamefont {{Mirrahimi}}}, \ and\ \bibinfo {author}
  {\bibfnamefont {M.~H.}\ \bibnamefont {{Devoret}}},\ }\bibfield  {title}
  {\enquote {\bibinfo {title} {{A stabilized logical quantum bit encoded in
  grid states of a superconducting cavity}},}\ }\href@noop {} {\bibfield
  {journal} {\bibinfo  {journal} {arXiv e-prints}\ ,\ \bibinfo {eid}
  {arXiv:1907.12487}} (\bibinfo {year} {2019})},\ \Eprint
  {http://arxiv.org/abs/1907.12487} {arXiv:1907.12487 [quant-ph]} \BibitemShut
  {NoStop}%
\bibitem [{\citenamefont {Gottesman}\ \emph {et~al.}(2001)\citenamefont
  {Gottesman}, \citenamefont {Kitaev},\ and\ \citenamefont
  {Preskill}}]{PhysRevA.64.012310}%
  \BibitemOpen
  \bibfield  {author} {\bibinfo {author} {\bibfnamefont {Daniel}\ \bibnamefont
  {Gottesman}}, \bibinfo {author} {\bibfnamefont {Alexei}\ \bibnamefont
  {Kitaev}}, \ and\ \bibinfo {author} {\bibfnamefont {John}\ \bibnamefont
  {Preskill}},\ }\bibfield  {title} {\enquote {\bibinfo {title} {Encoding a
  qubit in an oscillator},}\ }\href {\doibase 10.1103/PhysRevA.64.012310}
  {\bibfield  {journal} {\bibinfo  {journal} {Phys. Rev. A}\ }\textbf {\bibinfo
  {volume} {64}},\ \bibinfo {pages} {012310} (\bibinfo {year}
  {2001})}\BibitemShut {NoStop}%
\bibitem [{\citenamefont {Steane}(1997)}]{PhysRevLett.78.2252}%
  \BibitemOpen
  \bibfield  {author} {\bibinfo {author} {\bibfnamefont {A.~M.}\ \bibnamefont
  {Steane}},\ }\bibfield  {title} {\enquote {\bibinfo {title} {Active
  stabilization, quantum computation, and quantum state synthesis},}\ }\href
  {\doibase 10.1103/PhysRevLett.78.2252} {\bibfield  {journal} {\bibinfo
  {journal} {Phys. Rev. Lett.}\ }\textbf {\bibinfo {volume} {78}},\ \bibinfo
  {pages} {2252--2255} (\bibinfo {year} {1997})}\BibitemShut {NoStop}%
\bibitem [{\citenamefont {Glancy}\ and\ \citenamefont
  {Knill}(2006)}]{PhysRevA.73.012325}%
  \BibitemOpen
  \bibfield  {author} {\bibinfo {author} {\bibfnamefont {S.}~\bibnamefont
  {Glancy}}\ and\ \bibinfo {author} {\bibfnamefont {E.}~\bibnamefont {Knill}},\
  }\bibfield  {title} {\enquote {\bibinfo {title} {Error analysis for encoding
  a qubit in an oscillator},}\ }\href {\doibase 10.1103/PhysRevA.73.012325}
  {\bibfield  {journal} {\bibinfo  {journal} {Phys. Rev. A}\ }\textbf {\bibinfo
  {volume} {73}},\ \bibinfo {pages} {012325} (\bibinfo {year}
  {2006})}\BibitemShut {NoStop}%
\bibitem [{\citenamefont {{Albert}}\ \emph {et~al.}(2017)\citenamefont
  {{Albert}}, \citenamefont {{Noh}}, \citenamefont {{Duivenvoorden}},
  \citenamefont {{Young}}, \citenamefont {{Brierley}}, \citenamefont
  {{Reinhold}}, \citenamefont {{Vuillot}}, \citenamefont {{Li}}, \citenamefont
  {{Shen}}, \citenamefont {{Girvin}}, \citenamefont {{Terhal}},\ and\
  \citenamefont {{Jiang}}}]{2017arXiv170805010A}%
  \BibitemOpen
  \bibfield  {author} {\bibinfo {author} {\bibfnamefont {V.~V.}\ \bibnamefont
  {{Albert}}}, \bibinfo {author} {\bibfnamefont {K.}~\bibnamefont {{Noh}}},
  \bibinfo {author} {\bibfnamefont {K.}~\bibnamefont {{Duivenvoorden}}},
  \bibinfo {author} {\bibfnamefont {D.~J.}\ \bibnamefont {{Young}}}, \bibinfo
  {author} {\bibfnamefont {R.~T.}\ \bibnamefont {{Brierley}}}, \bibinfo
  {author} {\bibfnamefont {P.}~\bibnamefont {{Reinhold}}}, \bibinfo {author}
  {\bibfnamefont {C.}~\bibnamefont {{Vuillot}}}, \bibinfo {author}
  {\bibfnamefont {L.}~\bibnamefont {{Li}}}, \bibinfo {author} {\bibfnamefont
  {C.}~\bibnamefont {{Shen}}}, \bibinfo {author} {\bibfnamefont {S.~M.}\
  \bibnamefont {{Girvin}}}, \bibinfo {author} {\bibfnamefont {B.~M.}\
  \bibnamefont {{Terhal}}}, \ and\ \bibinfo {author} {\bibfnamefont
  {L.}~\bibnamefont {{Jiang}}},\ }\bibfield  {title} {\enquote {\bibinfo
  {title} {{Performance and structure of single-mode bosonic codes}},}\
  }\href@noop {} {\bibfield  {journal} {\bibinfo  {journal} {ArXiv e-prints}\ }
  (\bibinfo {year} {2017})},\ \Eprint {http://arxiv.org/abs/1708.05010}
  {arXiv:1708.05010 [quant-ph]} \BibitemShut {NoStop}%
\bibitem [{\citenamefont {{Noh}}\ \emph {et~al.}(2019)\citenamefont {{Noh}},
  \citenamefont {{Albert}},\ and\ \citenamefont {{Jiang}}}]{8482307}%
  \BibitemOpen
  \bibfield  {author} {\bibinfo {author} {\bibfnamefont {K.}~\bibnamefont
  {{Noh}}}, \bibinfo {author} {\bibfnamefont {V.~V.}\ \bibnamefont {{Albert}}},
  \ and\ \bibinfo {author} {\bibfnamefont {L.}~\bibnamefont {{Jiang}}},\
  }\bibfield  {title} {\enquote {\bibinfo {title} {Quantum capacity bounds of
  gaussian thermal loss channels and achievable rates with
  gottesman-kitaev-preskill codes},}\ }\href {\doibase
  10.1109/TIT.2018.2873764} {\bibfield  {journal} {\bibinfo  {journal} {IEEE
  Transactions on Information Theory}\ }\textbf {\bibinfo {volume} {65}},\
  \bibinfo {pages} {2563--2582} (\bibinfo {year} {2019})}\BibitemShut {NoStop}%
\bibitem [{\citenamefont {{Pantaleoni}}\ \emph {et~al.}(2019)\citenamefont
  {{Pantaleoni}}, \citenamefont {{Baragiola}},\ and\ \citenamefont
  {{Menicucci}}}]{2019arXiv190708210P}%
  \BibitemOpen
  \bibfield  {author} {\bibinfo {author} {\bibfnamefont {Giacomo}\ \bibnamefont
  {{Pantaleoni}}}, \bibinfo {author} {\bibfnamefont {Ben~Q.}\ \bibnamefont
  {{Baragiola}}}, \ and\ \bibinfo {author} {\bibfnamefont {Nicolas~C.}\
  \bibnamefont {{Menicucci}}},\ }\bibfield  {title} {\enquote {\bibinfo {title}
  {{Modular Bosonic Subsystem Codes}},}\ }\href@noop {} {\bibfield  {journal}
  {\bibinfo  {journal} {arXiv e-prints}\ ,\ \bibinfo {eid} {arXiv:1907.08210}}
  (\bibinfo {year} {2019})},\ \Eprint {http://arxiv.org/abs/1907.08210}
  {arXiv:1907.08210 [quant-ph]} \BibitemShut {NoStop}%
\bibitem [{\citenamefont {Vuillot}\ \emph {et~al.}(2019)\citenamefont
  {Vuillot}, \citenamefont {Asasi}, \citenamefont {Wang}, \citenamefont
  {Pryadko},\ and\ \citenamefont {Terhal}}]{PhysRevA.99.032344}%
  \BibitemOpen
  \bibfield  {author} {\bibinfo {author} {\bibfnamefont {Christophe}\
  \bibnamefont {Vuillot}}, \bibinfo {author} {\bibfnamefont {Hamed}\
  \bibnamefont {Asasi}}, \bibinfo {author} {\bibfnamefont {Yang}\ \bibnamefont
  {Wang}}, \bibinfo {author} {\bibfnamefont {Leonid~P.}\ \bibnamefont
  {Pryadko}}, \ and\ \bibinfo {author} {\bibfnamefont {Barbara~M.}\
  \bibnamefont {Terhal}},\ }\bibfield  {title} {\enquote {\bibinfo {title}
  {Quantum error correction with the toric gottesman-kitaev-preskill code},}\
  }\href {\doibase 10.1103/PhysRevA.99.032344} {\bibfield  {journal} {\bibinfo
  {journal} {Phys. Rev. A}\ }\textbf {\bibinfo {volume} {99}},\ \bibinfo
  {pages} {032344} (\bibinfo {year} {2019})}\BibitemShut {NoStop}%
\bibitem [{\citenamefont {{Tzitrin}}\ \emph {et~al.}(2019)\citenamefont
  {{Tzitrin}}, \citenamefont {{Bourassa}}, \citenamefont {{Menicucci}},\ and\
  \citenamefont {{Sabapathy}}}]{TBMSK2019}%
  \BibitemOpen
  \bibfield  {author} {\bibinfo {author} {\bibfnamefont {Ilan}\ \bibnamefont
  {{Tzitrin}}}, \bibinfo {author} {\bibfnamefont {J.~Eli}\ \bibnamefont
  {{Bourassa}}}, \bibinfo {author} {\bibfnamefont {Nicolas~C.}\ \bibnamefont
  {{Menicucci}}}, \ and\ \bibinfo {author} {\bibfnamefont {Krishna~Kumar}\
  \bibnamefont {{Sabapathy}}},\ }\bibfield  {title} {\enquote {\bibinfo {title}
  {{Towards practical qubit computation using approximate error-correcting grid
  states}},}\ }\href@noop {} {\bibfield  {journal} {\bibinfo  {journal} {arXiv
  e-prints}\ ,\ \bibinfo {eid} {arXiv:1910.03673}} (\bibinfo {year} {2019})},\
  \Eprint {http://arxiv.org/abs/1910.03673} {arXiv:1910.03673 [quant-ph]}
  \BibitemShut {NoStop}%
\bibitem [{\citenamefont {Gu}\ \emph {et~al.}(2009)\citenamefont {Gu},
  \citenamefont {Weedbrook}, \citenamefont {Menicucci}, \citenamefont {Ralph},\
  and\ \citenamefont {van Loock}}]{PhysRevA.79.062318}%
  \BibitemOpen
  \bibfield  {author} {\bibinfo {author} {\bibfnamefont {Mile}\ \bibnamefont
  {Gu}}, \bibinfo {author} {\bibfnamefont {Christian}\ \bibnamefont
  {Weedbrook}}, \bibinfo {author} {\bibfnamefont {Nicolas~C.}\ \bibnamefont
  {Menicucci}}, \bibinfo {author} {\bibfnamefont {Timothy~C.}\ \bibnamefont
  {Ralph}}, \ and\ \bibinfo {author} {\bibfnamefont {Peter}\ \bibnamefont {van
  Loock}},\ }\bibfield  {title} {\enquote {\bibinfo {title} {Quantum computing
  with continuous-variable clusters},}\ }\href {\doibase
  10.1103/PhysRevA.79.062318} {\bibfield  {journal} {\bibinfo  {journal} {Phys.
  Rev. A}\ }\textbf {\bibinfo {volume} {79}},\ \bibinfo {pages} {062318}
  (\bibinfo {year} {2009})}\BibitemShut {NoStop}%
\end{thebibliography}%
\appendix

\section{Notation and Definitions}
\label{appendix-defs}
Throughout our text, we use the following notations and conventions:

\begin{itemize}
\item Plank's constant $\hbar = 1$. 
\item Displacement operator $\hat{D}(\alpha) = \text{e}^{\frac{\alpha \hat{a}^{\dagger}-\alpha^{*} \hat{a}}{\sqrt{2}}}$.
\item Beam-splitter operator $\hat{B}(\phi) = \text{e}^{-\frac{\phi}{2}(\hat{a}^{\dagger}\hat{b}-\hat{a}\hat{b}^{\dagger})}$.
\item Squeezing operator $\hat{S}_{q}(a) = \text{e}^{\frac{\text{ln}(a)}{2}(\hat{a}^2-(\hat{a}^{\dagger})^2)}$.
\item Gaussian function $G_{\sigma}(x) = \exp(-\frac{x^2}{2 \sigma^2})$.
\item $\mathcal{N}(\mu,V)$ is a Gaussian random vector with mean $\mu$ and covariance $V$.
\item Rounding function $\nint*{x}$ gives the nearest integer to $x$.
\item Remainder function $\mathrm{rem}(n, 4)$ gives the remainder from dividing integer $n$ by 4.
\end{itemize}
Additionally, we define: 
\begin{itemize}
\item GKP wavefunctions widths $\vec{\Delta}=(\Delta,\kappa)$ and $\vec{\Delta'}=(\Delta/\sqrt{2},\kappa\sqrt{2})$.
\item Normalized GKP wavefunctions
\begin{equation}
   \psi_{\mu}^{\vec{\Delta}} (x) = N_\mu \sum_{s \in \mathbb{Z}} G_{\frac{1}{\kappa}} [(2s+\mu)\sqrt{\pi} \hspace{0.04cm}] \hspace{0.03cm} G_{\Delta}[x-(2s+\mu)\sqrt{\pi} \hspace{0.02cm}]. 
\end{equation}
\item Step functions
\begin{equation}
\begin{split}
        & s(x) = \frac{1}{2}\mathrm{rem}\Big{(}\nint*{x/\sqrt{\pi/2}} \ , \ 4 \Big{)} \ , \\
        & f^{*}_{\text{step}}(x) = \frac{\sqrt{\pi}}{2}\mathrm{rem}^{*}\Big{(}\nint*{x/\sqrt{\pi/2}} \ , \ 4 \Big{)} \ .
        \end{split}
\end{equation}
\item The effective measurement results $\mathcal{X}_{\text{m}}^{(h)}$ and effective errors $\mathcal{U}_{h}$ if passive error correction is carried out:
\begin{equation}
    \begin{split}
    & \mathcal{X}^{(h)}_{\text{m}} = x^{(h)}_{\text{m}} + \frac{1}{\sqrt{2}}\sum_{k=1}^{h-1}\Big[\frac{f_\text{step}^{*}(\mathcal{X}_\text{m}^{(k)})}{2^{h-k-1}}\Big] \ , \ \forall \ h > 1  \\
    & \mathcal{X}^{(1)}_{\text{m}} = x^{(1)}_{\text{m}} \\
    & \mathcal{U}_h = u_h + \sum_{k=1}^{h-1}\Big[\frac{u_k}{2^{h-k}} - \frac{f_\text{step}^{*}(\mathcal{X}_\text{m}^{(k)})}{2^{h-k-1}}\Big] \ , \ \forall \ h > 1 \\
    & \mathcal{U}_1 = u_1 \ . 
    \end{split}
    \label{eq:XU1}
\end{equation}
\end{itemize}

\section{Single-Round Error Estimation}
\label{appendix-single-step}
Here we present further details for a single round of syndrome extraction (SE) and error estimation. 

\subsection{Wavefunction after the q-SE circuit}
First, we show that the state after the q-SE circuit is approximated by simple transformation of its initial wavefunction. 
\begin{theorem}
\label{result_width_trick_q_QEC} A q-SE circuit  with measurement result $x_\mathrm{m}$ approximately transforms an input qubit wavefunction according to
\begin{equation*}
Q_{\alpha}^{\vec{\Delta}}(x-u)\rightarrow Q_{s(x_\mathrm{m})+\frac{\alpha}{2}}^{\vec{\Delta'}}\big(x-\frac{u}{2}\big).
\end{equation*}

\begin{proof} The two-mode input to the q-SE circuit has the wavefunction \vspace{0.2cm}
\begin{equation}\label{eq:psi_xy}
\Phi_\mathrm{in} (x,y) = Q_{\alpha}^{\vec{\Delta}}(x-u) \cdot \psi_{+}^{\vec{\Delta}}(y),
\end{equation}
where the qubit $Q_{\alpha}^{\vec{\Delta}}= a_{0}\psi_{0+\alpha}^{\vec{\Delta}}+a_{1}\psi_{1+\alpha}^{\vec{\Delta}}$ has an error shift $u$ and relative peak shift $\alpha$.

After the beam-splitter and squeezer of the q-SE circuit, the wavefunction is
\begin{equation}\label{eq:qSE_wavefunction1}
\Phi(x,y) = Q_{\alpha}^{\vec{\Delta}}(x+\frac{y}{\sqrt{2}}-u) \cdot \psi_{+}^{\vec{\Delta}}(-x+\frac{y}{\sqrt{2}}),
\end{equation}
which may be expanded as  
\begin{equation}
\begin{split}
\Phi(x,y)=  \sum_{\mu,n,m} & a_{\mu} N_{\mu} G_{\frac{1}{\kappa}} [(2n+\mu+\alpha)\sqrt{\pi}\hspace{0.04cm}] \\
& \cdot G_{\Delta}[x-u+\frac{y}{\sqrt{2}}-(2n+\mu+\alpha)\sqrt{\pi}\hspace{0.04cm}] \\
& \cdot G_{\frac{1}{\kappa}} (m\sqrt{\pi}\hspace{0.04cm}) \cdot \hspace{0.05cm} G_{\Delta}(-x+\frac{y}{\sqrt{2}}-m\sqrt{\pi} \hspace{0.04cm}),
\end{split} 
\end{equation}
where the sums are for integer $n,m$ and $\mu \in \{0,1\}$.

We now identify that this may be written in terms of un-rotated GKP wavefunctions. To do so, we regroup Gaussian functions of the same width by completing the square in the exponent, resulting in 

\begin{equation}\label{eq:qSE_wavefunction}
\begin{split}
\Phi(x,y) & \propto \sum_{\beta} \psi_{-\beta+\frac{\alpha}{2}}^{\vec{\Delta}'} \Big{(}\frac{y}{\sqrt{2}}-\frac{u}{2}\Big{)} \\
&\cdot \Big{(}a_0 \psi_{\beta+\frac{\alpha}{2}}^{\vec{\Delta}'} \big(x-\frac{u}{2}\big) + a_1 \psi_{1+\beta+\frac{\alpha}{2}}^{\vec{\Delta}'} \big(x-\frac{u}{2}\big)\Big{)}
\end{split}
\end{equation}
where the sum is over $\beta\in\{-\frac{1}{2},0,\frac{1}{2},1\}$. This wavefunction is a superposition of two-mode GKP states with relative peaks shifts differing by $\beta$. Note also that the initial error $u$ is now shared equally by the modes. To simplify the expression, we have used the approximation that the normalization constants are equal: $\frac{1}{N_0} \approx \frac{N_0}{N_1^2} \approx \frac{N_0}{N_{1/2} N_{-1/2}}$.

For a measurement result $x_\mathrm{m}$, the output wavefunction is proportional to $\Phi(x,x_\mathrm{m})$. For a particular $x_\mathrm{m}$, one term from equation~(\ref{eq:qSE_wavefunction}) is much larger than the others, since the weighting functions are positive-valued and nearly orthogonal if $\Delta$ is sufficiently smaller than $\sqrt{\pi}/2$. The dominant term is given by $\beta = s(x_\mathrm{m}-\frac{u}{\sqrt{2}}-\alpha\sqrt\frac{\pi}{2})$, and therefore we approximate that the output state as that given by the most likely case $u=0$. 

The average fidelity of this approximation is given by the average value of $|\psi_{s(x_\mathrm{m})+\frac{\alpha}{2}}^{\vec{\Delta}'}|^2$ taken over the joint probability distribution $\mathbb{P}(x_\mathrm{m},u)=\mathbb{P}(x_\mathrm{m}|u)\mathbb{P}(u)$ (these distributions are discussed below).
\end{proof}
\end{theorem}

A similar result is obtained for the p-quadrature syndrome extraction. In this case, however, the widths undergo the inverse transformation according to $(\Delta,\kappa)\rightarrow (\Delta\sqrt{2},\kappa/\sqrt{2})$. Appropriately designed q-SE and p-SE steps can therefore be used sequentially to prevent any overall change in the widths. In particular, we begin with a q-SE step using a qubit and auxiliary state of width $(\Delta,\kappa)$. The following p-SE step then uses an auxiliary state of width $(\Delta/\sqrt{2},\kappa/\sqrt{2})$, and the final state from this has the same width as the initial one. We note that the effect of width reduction through syndrome extraction cannot be observed in the $\Delta = 0$ unphysical states calculation.  

\subsection{Wavefunction after error, q-SE and p-SE}
We now extend the above calculation to include a subsequent p-SE step. In particular, we will show the transformation of the input state is well approximated by
\begin{equation}
\label{eq:step_result} 
Q^{\vec{\Delta}}(x-u)\text{e}^{ivx} \rightarrow Q^{\vec{\Delta}}\big(x-\theta(x_\mathrm{m},u)\big)\text{e}^{i\theta(p_\mathrm{m},v)x} \hspace{0.1cm} \,
\end{equation}
where the shift function $\theta$ is defined as
\begin{equation}\label{eq:theta}
\theta(y,b)=b/2-f^{*}_\mathrm{step}(y).
\end{equation}

The two-mode state after the p-SE beam-splitter and squeezer is given by
\begin{equation}
\begin{split}
    \Phi(x,z) = \hspace{0.1cm} & \text{e}^{i(\frac{x}{2}+\frac{z}{\sqrt{2}})v} \cdot \psi^{\vec{\Delta'}}_{0}\Big(\frac{x}{2}-\frac{z}{\sqrt{2}}\Big) \\
    & \cdot Q^{\vec{\Delta'}}\Big(\frac{x}{2}+\frac{z}{\sqrt{2}}-\frac{u}{2}+f^{*}_\mathrm{step}(x_\mathrm{m}-u/\sqrt{2})\Big) \ ,
    \end{split}
\end{equation}
\noindent where the quadrature $z$ corresponds to the mode to be measured. This p-quadrature measurement is achieved using the Fourier transform of the two-mode wavefunction
\begin{equation}
\begin{split}
    \mathcal{F}\big\{\Phi &(x,z)\big\}  = \hspace{0.1cm} \mathcal{F}_{y\rightarrow p-\frac{p_{\text{m}}}{\sqrt{2}}}\Big\{ \psi^{\vec{\Delta'}}_{0}(y)\Big\} \\ & \cdot 
    \mathcal{F}_{w\rightarrow p+\frac{p_{\text{m}}}{\sqrt{2}}}\Big{\{} \text{e}^{iwv} Q^{\vec{\Delta'}}\big{(} w-\frac{u}{2}+f^{*}_\text{step}(x_\mathrm{m}-u/\sqrt{2})\Big{)} \big{\}} .
    \end{split}
\end{equation}
where the transform is taken for variables $(x,z) \rightarrow (p,p_\text{m})$, and on the right hand side we use the change of variables $w = \frac{x}{2}+\frac{z}{\sqrt{2}}$ and $y = \frac{x}{2}-\frac{z}{\sqrt{2}}$. The shift property of the transform yields 
\begin{equation}
\begin{split}\label{eq:pSE_wavefunction}
    \tilde{\Phi}(p,p_\mathrm{m}) = &\tilde{\psi}^{\vec{\Delta'}}_{0}(p-\frac{p_\mathrm{m}}{\sqrt{2}})\tilde{Q}^{\vec{\Delta'}}_{0}(p+\frac{p_\mathrm{m}}{\sqrt{2}}-v) \\
    &\cdot \text{e}^{-ip(u/2-f^{*}_\text{step}(x_\mathrm{m}-u/\sqrt{2}))}
    \end{split}
\end{equation}
where we denote Fourier transforms as $\tilde{g}=\mathcal{F}\{g\}$.

Since the finite-energy GKP states are approximate eigenstates of the Fourier transform, the first two terms are similar to the product of two rotated GKP states, treated above. Using similar methods, we find the approximate relationship
\begin{equation}
\begin{split}
\tilde{\psi}^{\vec{\Delta'}}_{0}\Big(p-\frac{p_{\text{m}}}{\sqrt{2}}\Big) &\cdot 
    \tilde{Q}^{\vec{\Delta'}}\Big(p+\frac{p_{\text{m}}}{\sqrt{2}}-v\Big) \\
    &\propto \tilde{Q}^{\vec{\Delta}}\Big(p - \frac{v}{2}+f^{*}_\mathrm{step}(p_\mathrm{m}-v/\sqrt{2})\Big).
    \label{eq:reindex_calc}
\end{split}
\end{equation}
We now simplify the step function by noting that $f^{*}_\mathrm{step}(p_\mathrm{m}-v/\sqrt{2}) \approx f^{*}_\mathrm{step}(p_\mathrm{m})$. The success probability of this approximation will be discussed later. Finally, we take the inverse Fourier transform to find the desired result given by equation~(\ref{eq:step_result}).

\subsection{Bayesian estimation of the q-quadrature error}
In order to estimate the unknown shift error, we wish to determine its posterior probability distribution. To do so, we use the prior probability distribution, specified by the error model, and a likelihood function $\mathbb{P}(x_\text{m}|u)$ that we now calculate from the wavefunction.

\begin{theorem}
\label{result_proof_of_P_x_m_given_u} The probability of measurement result $x_\mathrm{m}$ from a q-SE circuit with initial state $\psi^{\vec{\Delta}}_{\alpha}(x - u)$ is 
\begin{equation*}
\mathbb{P}(x_\mathrm{m}|u) \propto \psi^{\vec{\Delta}}_{+}(\sqrt{2}x_\mathrm{m} - u).   
\end{equation*}

\begin{proof}
We start with the two-mode wavefunction given in equation~(\ref{eq:qSE_wavefunction}). The probability for measurement result $x_\mathrm{m}$ is 
\begin{equation}
\begin{split}
 \mathbb{P}(x_\text{m}|u) & = \int_{x\in \mathbb{R}} \big|\Phi(x,x_\mathrm{m})\big|^2 dx \\
 & \approx \sum_{\beta} \Big{|}\psi_{\beta}^{\vec{\Delta}'} \Big{(}\frac{x_\text{m}}{\sqrt{2}}-\frac{u}{2}\Big{)}\Big{|}^2, 
\end{split}
\end{equation} 
where in the second step we neglect overlap of the GKP wavefunctions to approximate $|\sum \psi_{\beta}|^2 \approx \sum |\psi_{\beta}|^2$.  

An alternate expression is obtained from the relationships
\begin{equation}
\label{eq:reindexed}
\begin{aligned}
\sum_{\beta \in \{0,1 \} }&\Big{|}\psi_{\beta}^{\vec{\Delta}'} \Big{(}\frac{x_\text{m}}{\sqrt{2}}-\frac{u}{2}\Big{)}\Big{|}^2 \propto  \psi_{0}^{\vec{\Delta}} (0)  \psi_{0}^{\vec{\Delta}} \Big{(}\sqrt{2}x_\text{m}-u\Big{)} \\
\sum_{\beta \in \{\pm \frac{1}{2}\} }&\Big{|}\psi_{\beta}^{\vec{\Delta}'} \Big{(}\frac{x_\text{m}}{\sqrt{2}}-\frac{u}{2}\Big{)}\Big{|}^2 \propto  \psi_{0}^{\vec{\Delta}} (0)  \psi_{1}^{\vec{\Delta}} \Big{(}\sqrt{2}x_\text{m}-u\Big{)}, \\
\end{aligned}
\end{equation} 
which may be shown by re-indexing akin to the derivation of equation~(\ref{eq:reindex_calc}). Addition of these two equations yields the desired result. 
\end{proof}
\end{theorem}

Using Bayes' theorem, the posterior probability distribution is
\begin{equation}
\begin{aligned}
\mathbb{P}(u|x_\text{m}) &\propto \mathbb{P}(x_\text{m}|u) \mathbb{P}(u) \\
&\propto \psi^{\vec{\Delta}}_{+}(\sqrt{2}x_\text{m} - u)\text{e}^{-\frac{u^2}{2\sigma_0^2}},
\end{aligned}
\label{eq:p_u_x_m}
\end{equation}
where in the last step we have explicitly written the Gaussian error model. The unknown shift can now be estimated from equation~(\ref{eq:p_u_x_m}) using an appropriate estimator, such as the mean of $\mathbb{P}(u|x_\text{m})$. For a given $x_\text{m}$, along with $\Delta \ll \sqrt{\pi}$ and $u \ll 1$, $\mathbb{P}(u|x_\text{m})$ can be well approximated by a normal distribution
\begin{equation}
    \mathbb{P}(u|x_\text{m}) \approx \mathcal{N}\left(\frac{\sqrt{2} \sigma_0^2 x_\text{m} - \sqrt{\pi} \sigma_0^2  \nint*{\frac{\sqrt{2}x_\text{m}}{\sqrt{\pi}}}}{\Delta^2 + \sigma_0^2}, \frac{\Delta^2\sigma_0^2}{\Delta^2 + \sigma_0^2}\right) \ .
\end{equation}
\noindent The approximation of the posterior as a normal distribution given reasonable parameter choices is an important step used below in multi-round calculations.

\subsection{Bayesian estimation of q- and p-quadrature error}
We now expand on the previous section to estimate an unknown shift in both quadratures.
\begin{theorem}
\label{result_proof_of_P_x_m_p_m_given_u_v} The probability of measurement results $x_\text{m}$ and $p_\text{m}$ from sequential q-SE and p-SE circuits with an initial state $\mathrm{e}^{ivx} \psi_{\alpha}^{\vec{\Delta}}(x-u)$ is 
\begin{equation*}
 \mathbb{P}(x_\mathrm{m},p_\mathrm{m}|u,v) \propto \psi^{\vec{\Delta}}_{+}(\sqrt{2}x_\text{m} - u) \cdot  \psi^{(2\Delta,\kappa/2)}_{+}(\sqrt{2}p_\text{m} - v) \ .
\label{eq:P_x_m_p_m_given_u_v}
\end{equation*}

\begin{proof}
Starting from the wavefunction in equation~(\ref{eq:qSE_wavefunction}), the probability of $p_\mathrm{m}$ is  
\begin{equation}\label{eq:qSE_p_of_x_1} 
\begin{split}
 \mathbb{P}(p_\text{m}|u,v,x_\mathrm{m}) &= \int_{p \in \mathbb{R}} \big{|} \tilde{\Phi}(p,p_\mathrm{m})\big{|}^2\ dp \\
 &\propto \psi^{(2\Delta,\kappa/2)}_{+}(\sqrt{2}p_\text{m} - v),
\end{split}
\end{equation}
where the second step follows by analogy to equations~(\ref{eq:qSE_wavefunction1}-\ref{eq:qSE_wavefunction}) and (\ref{eq:qSE_p_of_x_1}-\ref{eq:reindexed}). The desired result follows from the relationship of conditional probabilities $\mathbb{P}(x_\mathrm{m},p_\mathrm{m}|u,v) =\mathbb{P}(x_\mathrm{m}|u,v)\mathbb{P}(p_\mathrm{m}|u,v,x_\mathrm{m})$. 
\end{proof}
\end{theorem}
Note that $\mathbb{P}(p_\mathrm{m}|u,v,x_\mathrm{m})$ is independent of $x_\mathrm{m}$ and $u$. We can therefore write $\mathbb{P}(x_\mathrm{m},p_\mathrm{m}|u,v)=\mathbb{P}(x_\mathrm{m}|u)\mathbb{P}(p_\mathrm{m}|v)$ where the $\mathbb{P}(p_\mathrm{m}|v)$ differs from $\mathbb{P}(x_\mathrm{m}|u)$ according to the transformation $\Delta \rightarrow 2\Delta, \kappa \rightarrow \kappa/2$.

\section{Multi-Round Error Estimation}
\label{appendix-multi-step}
We now consider $M$ rounds of error followed by q-SE and p-SE circuits, with an aim to estimate the shift error on the final state.
\subsection*{Wavefunction after multiple rounds}
As shown for the single-round scenario, the wavefunction after each round is a shifted version of the input. An expression for the shift after each round is calculated iteratively. After round $h$, with a history of measurements $\vec{x}_\mathrm{m}=(x_\mathrm{m}^{(1)},\dotsc ,x_\mathrm{m}^{(h)})$ and shift errors $\vec{u}=(u_1,\dotsc ,u_h)$, the total q-shift is 
\begin{equation}\label{eq:theta_h}
    \theta_h\big(\vec{x}_\mathrm{m},\vec{u}\big)=\sum_{k=1}^{h} \frac{1}{2^{h-k}}\theta\big(\mathcal{X}_k(\vec{x}_\mathrm{m}),u_k\big),
\end{equation}
where $\theta$ is defined in equation~(\ref{eq:theta}) and
\begin{equation}
\mathcal{X}^{(k)}_{\text{m}}(\vec{x}_\mathrm{m}) = x^{(k)}_{\text{m}} + \frac{1}{\sqrt{2}}\sum_{j=1}^{k-1}\Big[\frac{f_\text{step}^{*}\big(\mathcal{X}_\text{m}^{(j)}(\vec{x}_\mathrm{m})\big)}{2^{k-j-1}}\Big]
\end{equation}
for $h>1$, and $\mathcal{X}^{(1)}_{\text{m}} = x^{(1)}_{\text{m}}$. The parameter $\mathcal{X}^{(k)}_{\text{m}}$ can be interpreted as the measurement outcome $x^{(k)}_{\text{m}}$ transformed to account for previous measurement-induced shifts.

Equation~(\ref{eq:theta_h}) can also be expanded as
\begin{equation}
\begin{split}
\theta_h\big(\vec{x}_\mathrm{m},\vec{u}\big) &= \sum_{j=1}^{h}\Big[\frac{u_k}{2^{h-j+1}}\Big] -
\sum_{k=1}^{h}\Big[\frac{f_\text{step}^{*}\big(\mathcal{X}_\text{m}^{(k)}(\vec{x}_\mathrm{m})\big)}{2^{h-k}}\Big] \\
&= \theta_h^{\mathrm{err}}(\vec{u})-\theta_h^{\mathrm{step}}(\vec{x}_\mathrm{m}),
\end{split}
\end{equation}
which identifies the total shift as comprised of a known measurement-induced shift $\theta_h^{\mathrm{step}}(\vec{x}_\mathrm{m})$ and the accumulation of unknown error $\theta_h^{\mathrm{err}}(\vec{u})$.
The aim of our error estimation after $M$ rounds is to use $\vec{x}_\mathrm{m}$ to estimate $\theta_M^{\mathrm{err}}(\vec{u})$. 

An analogous approach to the p-shifts leads to the wavefunction
\begin{equation}
\text{e}^{i\theta_h(\vec{p}_\mathrm{m},\vec{v})x}Q^{\vec{\Delta}}\big(x-\theta_h(\vec{x}_\mathrm{m},\vec{u})\big)
\label{eq:main}
\end{equation}
after $h$ rounds, for an initial qubit state $Q^{\vec{\Delta}}(x)$. This is assuming that $f_{\text{step}}$ is independent of $u_{k}$, which is true if $f_\text{step}^{*}(x_\text{m}) = f_\text{step}^{*}(x_\text{m}-u/\sqrt{2})$. We shall discuss the error of such an approximation in a later section.

\subsection{Total qubit drift}
\noindent Without active corrective shift at each round of QEC, the magnitude of the total drift of the state in phase space can be bounded.
Using the fact that $|f_\text{step}^{*}| \leq \sqrt{\pi}$ and taking the variance of each the error shifts to be $\sigma_0^2$, an application of the triangle inequality gives:

\begin{equation}
\begin{split}
    |\theta(x_\text{m}^{(h)},\mathcal{U}_h)| \hspace{0.1cm} , \hspace{0.1cm} |\theta(p_\text{m}^{(h)},\mathcal{V}_h)| \leq 2\sqrt{\pi} & (1-2^{-h}) \\ & + \Big|\mathcal{N}\Big(0,\sigma_0^2\frac{(1-4^{-h})}{3}\Big)\Big| \ , 
    \end{split}
\end{equation}
\noindent where $\mathcal{N}(\mu,V)$ is a Gaussian random variable with mean $\mu$ and Variance $V$. 
The quantities $1-4^{-h}$ and $1-2^{-h}$ converge very quickly to 1 as $h$ increases, meaning that for multiple rounds we have
\begin{equation}
    |\theta(x_\text{m}^{(h)},\mathcal{U}_h)| \hspace{0.1cm} , \hspace{0.1cm} |\theta(p_\text{m}^{(h)},\mathcal{V}_h)| \leq 2\sqrt{\pi} + \Big|\mathcal{N}\Big(0,\frac{\sigma_0^2}{3}\Big)\Big| \ .
\end{equation}
\noindent This means that even without active error corrective shift after each round of syndrome extraction, the GKP qubit is expected to drift by a maximum of $2\sqrt{\pi}$ in phase space along with some Gaussian random variable with variance $\sigma_0^2/3$ in the worst case scenario. 
Consequently, the energy of the physical system is not divergent if we extend the QEC for many rounds, if errors are sufficiently small at each round.

\subsection{Memory-assisted decoder}
\noindent The probability distributions for the $x$ and $p$ measurement values obtained in the $h^{\text{th}}$ round, given all the previous measurement results and displacement shifts, are given by expressions similar in spirit to that in Theorem~\ref{result_proof_of_P_x_m_given_u}:

\begin{equation}
    \begin{split}
    &\mathbb{P}(x^{(h)}_{\text{m}}|u_1,...,u_h,x^{(1)}_{\text{m}},...,x^{(h-1)}_{\text{m}}) \propto \psi^{(\Delta,\Delta)}_{+}(\sqrt{2}x^{(h)}_{\text{m}}-\mathcal{U}_h)\\
    & \mathbb{P}(p^{(h)}_{\text{m}}|v_1,...,v_h,p^{(1)}_{\text{m}},...,p^{(h-1)}_{\text{m}}) \propto \psi^{(2\Delta,\Delta/2)}_{+}(\sqrt{2}p^{(h)}_{\text{m}}-\mathcal{V}_h).
    \end{split}
\end{equation}

\noindent Using Bayes' Theorem to flip these distributions, for $M$ rounds we obtain
    \begin{equation}
\begin{split}
\mathbb{P}_{M}(\vec{u},\vec{v}|\vec{x}_\text{m},\vec{p}_\text{m}) = \mathbb{P}^{(\text{q})}_{M}(\vec{u}|\vec{x}_\text{m}) \cdot \hspace{0.05cm} \mathbb{P}^{(\text{p})}_{M}(\vec{v}|\vec{p}_\text{m}) 
\\ 
\mathbb{P}^{(\text{q})}_{M}(\vec{u}|\vec{x}_\text{m}) \propto \prod_{h=1}^{M}\psi^{\vec{\Delta}}_{+}(\sqrt{2}x^{(h)}_{\text{m}}-\mathcal{U}_h) \cdot \hspace{0.05cm} G_{\sigma_0}(u_h)
\\
\mathbb{P}^{(\text{p})}_{M}(\vec{v}|\vec{p}_\text{m}) \propto \prod_{h=1}^{M}\psi^{(2\Delta,\Delta/2)}_{+}(\sqrt{2}p^{(h)}_{\text{m}}-\mathcal{V}_h) \cdot \hspace{0.05cm} G_{\sigma_0}(v_h).
\end{split}
\end{equation}

\noindent We can see that the q- and p-quadratures equations have the similar forms and can be unchangeable ($q\leftrightarrow p$) if $u\leftrightarrow v$, $x_{\text{m}}\leftrightarrow p_\text{m}$ and $(\Delta,\kappa) \leftrightarrow (2\Delta,\kappa/2)$, so we will focus on the q-quadature version for now. The PDF for $\vec{u}$ is well approximated by a multivariate Gaussian distribution, a step known as Laplace's approximation in statistics, which we write as

\begin{equation}
\mathbb{P}^{(\text{q})}_{M}(\vec{u}|\vec{x}_\text{m}) \approx \mathcal{N}(\vec{\tilde{u}} , \Sigma) \ ,
\label{eq:gauss_approx}
\end{equation}
for mean vector $\vec{\tilde{u}}$ and covariance matrix $\Sigma$. 

We derive the inverse of the covariance matrix, $\Sigma^{-1}$, exactly by completing the square of the exponents of the Gaussians in equation~\eqref{eq:gauss_approx}:

\begin{equation}
\begin{split}
   & (\Sigma^{-1})_{\alpha,\beta} = \Big(\frac{\delta_{\alpha,\beta}}{\sigma_0^2} + \frac{1}{\Delta^2}{\displaystyle\sum_{h=m_{\alpha,\beta}}^{M}\frac{1}{4^{h}2^{-(\alpha+\beta)}}}\Big) \ , \\
   & m_{\alpha,\beta} = \max{\{\alpha,\beta\}} \ .
   \end{split}
\end{equation}

To derive $\vec{\tilde{u}}$ via completing the square, we need to invert the matrix $\Sigma^{-1}$. It's generally hard to invert a $M\times M$ symmetric matrix for an arbitrary M, but we can approximate the covariance matrix via a Neumann series expansion of the matrices, under the assumption of small displacements. We find the covariance matrix has components

\begin{equation}
    \Sigma_{\alpha,\beta} \approx \sigma_0^2 \hspace{0.05cm} \delta_{\alpha,\beta} -  \frac{\sigma_0^4}{\Delta^2}  \sum_{h=\text{m}_{\alpha,\beta}}^{M}\frac{1}{4^{h}2^{-(\alpha+\beta)}} + \mathcal{O}\Big(\frac{\sigma_0^6}{\Delta^4}\Big) \ , 
\end{equation}
\noindent and note that this Neumann series converges if $\frac{\sigma_0}{\Delta} < \frac{1}{2}$ (which works for all $M$). By ignoring from order $\mathcal{O}\Big(\frac{\sigma_0^6}{\Delta^6}\Big)$ upwards in the expansion, we compute

\begin{equation}
\begin{split}
    \tilde{u}_{k} = \Big(\frac{\sigma_0}{\Delta}\Big)^2 \Bigg\{&2^{k} \sum_{j=1}^{M}\frac{F_j}{2^j} \\
    & - \Big(\frac{\sigma_0}{\Delta}\Big)^2\sum_{h=1}^{M}\Big[\Big(\hspace{-0.2cm}\sum_{n=\text{m}_{k,h}}^{M}\hspace{-0.2cm}\frac{2^{k+h}}{4^n}\Big)\Big(\sum_{j=h}^{M}\frac{F_j}{2^j}\Big)\Big]2^{h}\Bigg\} \ ,
\end{split}
\end{equation}
\noindent where $\text{m}_{k,h} = \text{max}\{k,h\}$ and $F_h = \sqrt{2}\mathcal{X}_\text{m}^{(h)}-\sqrt{\pi}\nint*{\sqrt{2}\mathcal{X}_\text{m}^{(h)}/\sqrt{\pi}}$, which is equation \eqref{eq:u_tilde_k} in the main text.

Now that we have characterised the multivariate Gaussian noise model (multivariate due to many rounds) by its covariance matrix, we wish to estimate the random part of the total displacement after $M$ rounds, given by $\sum_{k=1}^M u_k/2^{M-k+1}$. The corresponding PDF is 

\begin{equation}
\mathbb{P}\Big(\sum_{k=1}^{M}\frac{ u_k}{2^{M-k+1}}\Big|\vec{x}_{\text{m}}\Big) = \mathcal{N}(\vec{a}\cdot\vec{\tilde{u}} \ , \ \vec{a}^{\hspace{0.05cm}\text{T}}\hspace{-0.05cm} 
   \cdot\Sigma\cdot
    \vec{a}) \ ,
\end{equation}
\noindent with $a_{k} = 2^{-(M+1-k)}$.

We use the mean of this posterior distribution as the MMSE estimator, with uncertainty given by the variance 
\begin{equation}
   V_\text{q}=\text{Var}\Big(\sum_{k=1}^{M}\frac{u_k}{2^{M-k+1}}\Big) = \vec{a}^{\hspace{0.05cm}\text{T}}\hspace{-0.05cm} 
   \cdot\Sigma\cdot
    \vec{a}.
\end{equation}

\noindent Neglecting terms of order $\big(\frac{\sigma_0^6}{\Delta^4}\big)$ and higher, we compute 
\begin{equation}
\begin{split}
    \label{eq:var_q_multiple_steps2}
V_\text{q}(M) = & \frac{\sigma_{0}^2}{3}\Big[(1-4^{-M}) + \\ 
& \Big(\frac{2\sigma_0}{3\Delta}\Big)^{2}(4^{-2M} + 3(1+2M)4^{-M} -4)\Big] \ .
\end{split}
\end{equation}
\noindent We can see that $V_\text{q} \rightarrow \frac{\sigma_{0}^2}{3}$, very quickly as $M$ grows. Note that using measurement results from all rounds of syndrome measurements allows the error to be estimated more precisely than is achievable otherwise. 
\section{Approximations and qubit fidelity}
\label{appendix-success-prob}
In our calculations, two significant approximations are made in expressing the qubit wavefunction emerging from a syndrome extraction circuit as a displacement of the incoming wavefunction. For the single-round case, these were discussed following equation~\eqref{eq:qSE_wavefunction}. First, in identifying the dominant GKP-like term in the resulting wavefunction, we approximate that $f_\text{step}^{*}(x_\text{m}-u/\sqrt{2})=f_\text{step}^{*}(x_\text{m})$. Second, we keep only the dominant term in the sum of equation~\eqref{eq:qSE_wavefunction}. Each of these approximations reduce the fidelity of our description of the final state.

\subsection{Tracking error}
Unlikely measurement outcomes and unusually large displacement errors can cause an incorrect identification of the dominant GKP-like term, which we refer to as a tracking error. In this section, we estimate the probability that our description of the state after one or multiple rounds of syndrome extraction, as given in equation~\ref{eq:main}, does not contain a tracking error.

\subsubsection{Single round}
For a single round, the joint probability density function of $x_\text{m}$ and $u$ is
\begin{equation}
    \mathbb{P}(u,x_\text{m}) = C\sum_{n\in\mathbb{Z}}G_{1/\Delta}(n\sqrt{\pi})G_{\Delta}(\sqrt{2}x_\text{m}-u-n\sqrt{\pi}) G_{\sigma_0}(u) \ .
\end{equation}
No tracking error occurs if $x_\text{m}$ and $u$ are in the region $\mathcal{S}$ for which $|f_\text{step}^{*}(x_\text{m}) - f_\text{step}^{*}(x_\text{m}-u/\sqrt{2})| = 0$.

We therefore define the success probability for one round and error width $\sigma_0$ to be
\begin{equation}
    P^{\text{succ}}_{\text{track}}(1,\sigma_0) = \int_{S}  \mathbb{P}(u,x_\text{m}) \text{d}\mathcal{S} \ ,
\end{equation}
for which the integration region could be re-written as $x_\text{m}\in\mathbb{R}$ and $u\in\mathcal{T}$, where 
 $\mathcal{T} = [-\sqrt{\pi},0]+\sqrt{2}\text{mod}(x-\sqrt{\frac{\pi}{2}}/2,\sqrt{\frac{\pi}{2}})$. The normalisation constant $C = [\sqrt{2}\pi\sigma_0\Delta\Theta_3(0,\text{e}^{-\pi\Delta^2/2})]^{-1}$, where $\Theta_3$ is the Jacobi-Theta function of the third kind. 
 
 For our example in the main text where $(\Delta,\sigma_0^2) = (0.2182,0.0005)$, we calculate $1-P^{\text{succ}}_{\text{track}}(1,\sigma) \approx 1\times10^{-5}$.
\subsubsection{Multiple rounds}
For multiple rounds, the joint probability density function of $\vec{x}_{\text{m}}$ and $\vec{u}$ is 
\begin{equation}
    \mathbb{P}(\vec{u},\vec{x}_\text{m}) = \prod_{h=1}^{M}\psi^{\vec{\Delta}}_{+}(\sqrt{2}x^{(h)}_{\text{m}}-\mathcal{U}_h) \cdot \hspace{0.05cm} G_{\sigma_0}(u_h)
    \label{eq:pdf_joint_multi}
\end{equation}

Similar to the single round case, the probability for $M$ successful rounds is given by 
\begin{equation}
    P^{\text{succ}}_{\text{track}}(M,\sigma_0) = \int_{\mathcal{L}}  \mathbb{P}(\vec{u},\vec{x}_\text{m}) \text{d}\mathcal{L} \ ,
\end{equation}
where $\mathcal{L}$ is the region defined by $f^{*}_{\text{step}}(\mathcal{X}^{(j)}_\text{m}) = f^{*}_{\text{step}}(x^{(j)}_\text{m}-\mathcal{U}_j/\sqrt{2})$, $\forall 1<j<M$.

This calculation is simplified by changing variables to $\vec{\mathcal{X}}$ and $\vec{\mathbb{U}}$, where $\displaystyle\mathbb{U}_h = \sum_{k=1}\frac{u_k}{2^{h-k}}$ for $h>1$ and $\mathbb{U}_1 = u_1$, which leads to 
\begin{equation}
\begin{split}
       P^{\text{succ}}_{\text{track}}(M, & \sigma_0) >  \int_{\mathcal{L}}   \prod_{h=2}^{M}\Big[ \psi^{\vec{\Delta}}_{+}(\sqrt{2}\mathcal{X}^{(h)}_{\text{m}}-\mathbb{U}_h) \hspace{0.05cm} 
      \cdot G_{\frac{2\sigma_0}{\sqrt{5}}}(\mathbb{U}_h)\Big]
       \\ & 
      \cdot \psi^{\vec{\Delta}}_{+}(\sqrt{2}\mathcal{X}^{(1)}_{\text{m}}-\mathbb{U}_1) \cdot \hspace{0.05cm} G_{\sigma_0}(\mathbb{U}_h)
       \cdot \text{d}^{M}\hspace{-0.05cm}\vec{\mathcal{X}} \hspace{0.05cm}\text{d}^{M}\vec{\mathbb{U}} \ .
       \end{split}
\end{equation}
Note the multidimensional integral is now separated into a product of single integrals over the integration regions $f^{*}_{\text{step}}(\mathcal{X}^{(j)}_\text{m}) = f^{*}_{\text{step}}({\mathcal{X}^{(j)}_\text{m}-\mathbb{U}_j/\sqrt{2}})$ for all $2<j<M$. A convenient bound lower bound can be written in terms of single-round probabilities
\begin{equation}
\label{eq:multitrack}
           P^{\text{succ}}_{\text{track}}(M,\sigma_0) > \Big[P^{\text{succ}}_{\text{track}}\Big(1,\frac{2\sigma_0}{\sqrt{5}}\Big)\Big]^{M}.
\end{equation}
\noindent For $(\Delta,\sigma_0^2) = (0.2182,0.0005)$ and $M=200$, we calculate $1-P^{\text{succ}}_{\text{track}}(M,\sigma_0) \approx 3\times10^{-3}$.  

\subsection{Truncation error}
The approximation that only the dominant term in the state survives at each round of syndrome extraction leads to a probability that our decoder fails. We use an approximation of this probability to bound the long term fidelity of a state recovered using information from the memory-assisted decoder. In particular, we approximate the state after each round of syndrome extraction by decohering the dominant term with respect to the rest of the state: $p\hat{\rho}_\text{dom.} + (1-p)\hat{\rho}_\text{junk}$, where $p$ is the phenomenological success probability per round, $\hat{\rho}_\text{dom.}$ is the dominant term we keep and $\hat{\rho}_\text{junk}$ is the state we throw away in the truncation.

\subsubsection{Single round}
The wavefunction emerging from single syndrome extraction circuit, described by equation~\ref{eq:qSE_wavefunction}, can be expressed in ket notation as
\begin{equation}
\ket{\Phi(x|u,x_\text{m})} = R(u,x_\text{m}) \sum_{\gamma} \psi_{\gamma}^{\vec{\Delta}'} \Big{(}\frac{x_\text{m}}{\sqrt{2}}-\frac{u}{2}\Big{)} \ket{Q_\gamma^{\vec{\Delta}'}} \ ,
\end{equation}
\noindent where $R(u,x_\text{m})$ is a normalisation constant. We note the dominant term as $\ket{Q_\beta^{\vec{\Delta}}}$ and calculate its amplitude 

\begin{equation}
    A(\beta) = R(u,x_\text{m})\sum_{\gamma} \psi_{\gamma}^{\vec{\Delta}'} \Big(\frac{x_\text{m}}{\sqrt{2}}-\frac{u}{2}\Big) \braket{Q_{\beta}^{\vec{\Delta}}|Q_{\gamma}^{\vec{\Delta}'}} \ .
\end{equation}

The average success of the truncated description is given by 
\begin{equation}
    P^{\text{succ}}_{\text{trunc}}(1,\sigma_0)  = \int_{\text{all\ space}}  \hspace{-0.8cm}|A(\beta)|^2 \mathbb{P}(u,x_\text{m}) \ \text{d}u \hspace{0.05cm} \text{d}x_\text{m}.
\end{equation}

For the example $(\Delta,\sigma_0^2) = (0.2182,0.0005)$, we calculate $1-P^{\text{succ}}_{\text{trunc}}(1,\sigma_0) \approx 5\times 10^{-5}$.

\subsubsection{Multiple rounds}
Similarly, we estimate a success probability after $M$ rounds as
\begin{equation}
\begin{split}
    P^{\text{succ}}_{\text{trunc}}(M,\sigma_0)  = & \prod_{h=1}^{M}\int \sum_{\beta_h} |A(\beta_h)|^2 
    G_{\sigma_0}(u_h) \hspace{0.05cm} 
    \\ & \cdot \psi^{\vec{\Delta}}_{+}(\sqrt{2}x^{(h)}_{\text{m}}-\mathcal{U}_h) \ \text{d}\vec{u} \hspace{0.05cm} \text{d}\vec{x}_\text{m} \ .
\end{split}
\end{equation}
As in equation~\ref{eq:multitrack}, we bound the multi-round probability in terms of the single-round expression
\begin{equation}
\label{eq:multinorm}
P^{\text{succ}}_{\text{trunc}}(M,\sigma_0) > (P^{\text{succ}}_{\text{trunc}}(1,2\sigma_0/\sqrt{5}))^M \ .
\end{equation}
For $(\Delta,\sigma_0^2) = (0.2182,0.0005)$ and $M=200$ we calculate $1-P^{\text{succ}}_{\text{trunc}}(M,\sigma_0) \approx 1\times10^{-2}$.

\subsection{Total error and state fidelity}
To estimate a final qubit fidelity, as in the example shown in Fig.~\ref{fig:n_10_multiple_step}, we first calculate the fidelity $F_\rho$ from the output state described by our Bayesian estimation procedure, following the approach of Pantaleoni et al.~\cite{2019arXiv190708210P}. We then assume that a tracking, truncation or syndrome error results in a complete depolarized qubit with probability upper-bounded by $1-P^{\text{succ}}_\text{total}$. A lower bound on the final qubit fidelity is therefore given by $F = (F_\rho-\frac{1}{2})P^{\text{succ}}+\frac{1}{2}$.


\begin{algorithm}[!b]
\begin{algorithmic}
\Statex\textbf{Initialise} empty vectors $\vec{\mathcal{X}}_\text{m},\vec{F}, \vec{\tilde{u}}$
\\
    \FOR{h = 1 to M}
        \Statex \qquad Receive data $x_{\text{m}}^{(h)}$
        \\
        \Statex \qquad Compute $\displaystyle \mathcal{X}_\text{m}^{(h)} = x_\text{m}^{(h)} + \frac{1}{\sqrt{2}}\sum_{k=1}^{h-1}\frac{f^{*}_{\text{step}}(\mathcal{X}_\text{m}^{(k)})}{2^{h-k-1}}$
        \Statex \qquad
        \\
        \Statex \qquad Compute $\displaystyle F_h = \sqrt{2}\mathcal{X}_\text{m}^{(h)}-\sqrt{\pi}\nint*{\frac{\sqrt{2}\mathcal{X}_\text{m}^{(h)}}{\sqrt{\pi}}}$
        \\
        \Statex \qquad Append $\mathcal{X}_{\text{m}}^{(h)}$ to $\vec{\mathcal{X}}_\text{m}$ and $F_h$ to $\vec{F}$
    \ENDFOR
    
    \Statex Compute total displacement $\displaystyle \sum_{k=1}^{M} \frac{\tilde{u}_k}{2^{M-k+1}}$
    where
    \\
    \begin{equation*}
    \begin{split}
    \tilde{u}_{k} = \Big(\frac{\sigma_0}{\Delta}\Big)^2 \Bigg\{& 2^{k}\sum_{j=1}^{M}\frac{F_j}{2^j} \\ 
    & - \Big(\frac{\sigma_0}{\Delta}\Big)^2\sum_{h=1}^{M}\Big[\Big(\hspace{-0.2cm}\sum_{n=\text{m}_{k,h}}^{M}\hspace{-0.2cm}\frac{2^{k+h}}{4^n}\Big)\Big(\sum_{j=h}^{M}\frac{F_j}{2^j}\Big)\Big]2^{h}\Bigg\} \  \end{split}
    \end{equation*}
\end{algorithmic}
\caption{Estimation of the q-quadrature error}
\label{algor:1}
\end{algorithm}
\section{Pseudocode for Error Estimation}
\label{appendix-algorithm} After $M$ rounds of syndrome extraction, as depicted in Fig.~\ref{fig:memory}, our results can be used to estimated the random contribution of the total displacement error from measurement results $\vec{x}_\text{m}$ and $\vec{p}_\text{m}$. Algorithm~\ref{algor:1} gives an explicit description of this calculation for the q-quadrature.

\section{Offline Squeezing Derivation}
\label{appendix-offline-squeezing}
\begin{figure*}
\center
	\includegraphics[width=0.9\linewidth]{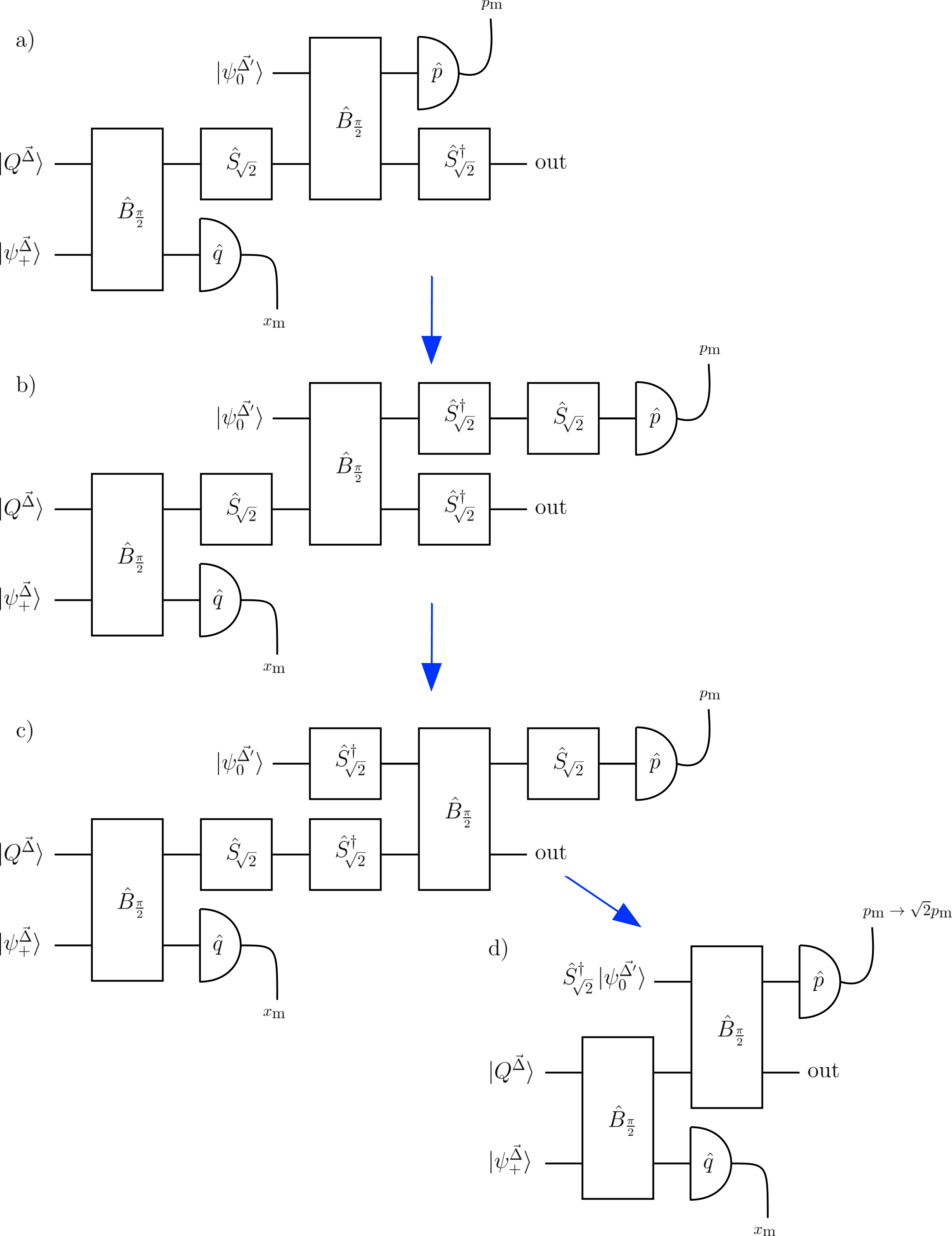}
	\caption{\label{fig:offline_sq_proof} Diagrammatic derivation of the offline squeezing circuit. In a), we start with the GK syndrome-extraction circuit with the corrective displacement removed. In b), we insert an identity operator before the momentum measurement. In c), we rearrange the 50:50 beam-splitter and squeezers in both modes. Finally in d), we achieve a simplified circuit in which all squeezing is effectively moved to the momentum resource state and interpretation of the momentum measurement result.}
\end{figure*}

A diagrammatic derivation of the offline squeezing circuit is shown in FIG.~\ref{fig:offline_sq_proof}. Starting with the Glancy-Knill syndrome extraction circuits, we recompile and re-interpret measurement results to obtain a circuit in which the qubit state interacts only with 50:50 beamsplitters.

\end{document}